\documentclass[preprint,preprintnumbers,amsmath,amssymb,floatfix,nofootinbib,superscriptaddress,longbibliography]{revtex4-1}

\usepackage{amsmath}
\usepackage{amssymb}
\usepackage[dvips]{graphicx}
\usepackage{url}
\usepackage{xcolor}
\usepackage[colorlinks,citecolor=blue,urlcolor=blue,linkcolor=blue]{hyperref}
\usepackage{booktabs} 

\usepackage[normalem]{ulem}

\newcommand{\hspazio}{~~~~~}
\newcommand{\spazio}{~~~~~~~~~~}

\begin{document}

\title{\boldmath $t\bar{t}b\bar{b}$ at the LHC: On the size of
  off-shell effects and prompt $b$-jet identification}

\preprint{TTK-22-06,  P3H-22-012, CAVENDISH-HEP-22/02}
\author{G. Bevilacqua}
\affiliation{ELKH-DE Particle Physics Research Group, University of Debrecen, 
H-4010 Debrecen, PBox 105, Hungary}
\author{H. Y. Bi}
\affiliation{Institute for Theoretical Particle Physics and Cosmology, 
RWTH Aachen University, D-52056 Aachen, Germany}
\author{H. B. Hartanto}
\affiliation{Cavendish Laboratory, University of Cambridge, J.J. Thomson Avenue, 
Cambridge CB3 0HE, United Kingdom}
\author{M. Kraus}
\affiliation{Physics Department, Florida State University, Tallahassee, 
FL 32306-4350, U.S.A.}
\author{M. Lupattelli}
\affiliation{Institute for Theoretical Particle Physics and Cosmology, 
RWTH Aachen University, D-52056 Aachen, Germany}
\author{M. Worek}
\affiliation{Institute for Theoretical Particle Physics and Cosmology, 
RWTH Aachen University, D-52056 Aachen, Germany}

\date{\today}

\begin{abstract}
We investigate full off-shell effects in $t\bar{t}b\bar{b}$
production in the dilepton channel at the LHC with the center-of-mass
energy $\sqrt{s} = 13$ TeV.  Specifically, we compute NLO QCD
corrections to the $pp \to e^+ \nu_e \mu^- \bar{\nu}_\mu b \bar{b} b
\bar{b} + X$ process and provide a prescription for $b$-jet identification to
distinguish prompt $b$ jets from $b$ jets originating from the decay
of the top quarks.  As an important irreducible background to $pp \to
t\bar{t}H (H\to b\bar{b})$, $t\bar{t}$ production in association with
two prompt $b$ jets is a primary source of uncertainty in the
measurement of $t\bar{t}H (H\to b\bar{b})$.  In quantifying full
off-shell effects, we perform comparisons between the state-of-the-art
full off-shell computation and the calculation in the
narrow width approximation.  The former includes all double-, single-
and non-resonant Feynman diagrams, interferences as well as
finite-width effects of the top quarks and $W$ gauge bosons. The
latter restricts the unstable top quarks and $W$ gauge bosons to
on-shell states and includes for the first time NLO QCD corrections to
both production and decays.  We observe that full off-shell effects are
subdominant compared to the scale uncertainties for the integrated
fiducial cross section and for the majority of differential
observables in the phase-space regions that we investigated.  However,
for a number of observables related to beyond the Standard Model
searches, full off-shell effects are significant.  Furthermore, with
our $b$-jet labelling  prescription, the prompt $b$ jets and the $b$
jets from top-quark decays can be successfully disentangled.

\end{abstract}

\maketitle


\section{Introduction}
\label{sec:intro}


The discovery of the Higgs boson at the Large Hadron Collider
(LHC)~\cite{ATLAS:2012yve, CMS:2012qbp} is one of the triumphs of the
Standard Model (SM). The Higgs boson plays an important role in the SM
since it is responsible for the electroweak (EW) symmetry breaking and
the origin of the masses of the elementary particles
\cite{Glashow:1961tr, Weinberg:1967tq, Salam:1968rm, Higgs:1964pj,
Englert:1964et}. The properties of the Higgs boson have been widely
tested since its discovery. Until now, this particle has proven to be
consistent with the expectations from the SM. The study of the
coupling of the Higgs boson to the heaviest of the fundamental
particles, the top quark, is crucial to test the SM and look for
effects of new physics.  The top-Yukawa coupling, denoted as $Y_t$,
could be probed, for instance, via Higgs boson production in the gluon
fusion process, in which the top quark appears in the loop. This $gg
\to H$ channel is also sensitive to possible beyond the Standard Model
(BSM) contributions, since new particles heavier than the top quark
can appear in the loop. A direct probe of $Y_t$ is provided by the
Higgs boson production in association with a top-quark pair,
$t\bar{t}H$ production, which was first observed at the LHC in
2018~\cite{ATLAS:2018mme,CMS:2018uxb}. Despite it being only 1\% of
the total Higgs boson production rate, the top-Higgs coupling is
already present at tree level.  The Higgs boson predominantly decays
into a bottom-quark pair $H\rightarrow b\bar{b}$ with a branching
ratio of $\mathcal{BR}=58\%$~\cite{ATLAS:2016neq}. Thus, $pp\to
t\bar{t}H$ with $H\to b\bar{b}$ can be measured with the best
statistical precision.  The measurement of $pp\to t\bar{t}H(H\to
b\bar{b})$ is very challenging due to the so-called combinatorial
background from the four $b$ jets in the final state and first
measurements have been reported in Ref.~\cite{ATLAS:2021qou}.  In
addition, the QCD process $pp \rightarrow t\bar{t}b\bar{b}$ has the
very same final state and represents an irreducible background to
$pp\to t\bar{t}H(H\to b\bar{b})$. The understanding of
$t\bar{t}b\bar{b}$ with higher precision could help us to better
isolate the Higgs boson signal from the background and, thus, improve
the measurement of $t\bar{t}H(H\to b\bar{b})$ production. Moreover,
$t\bar{t}b\bar{b}$ production is also important in some BSM physics
studies. For instance, ATLAS and CMS experiments conduct searches for
$\bar{t}bH^+(H^+\to t\bar{b})$~\cite{ATLAS:2021upq,CMS:2019rlz}, with
$t\bar{t}b\bar{b}$ as a background also to this process. However, no
significant excess above the expected SM background has been found so
far for this process.  Last but not least, $t\bar{t}b\bar{b}$ is
interesting on its own as it can provide an important test of QCD and
improve our understanding of the dynamics of the $g \to b\bar{b}$
splitting.

Precise predictions for $t\bar{t}b\bar{b}$ production were first
computed at NLO in QCD for stable top
quarks~\cite{Bredenstein:2008zb,Bredenstein:2009aj,
Bevilacqua:2009zn,Bredenstein:2010rs,Worek:2011rd}.
It was found that the NLO theoretical uncertainties stemming from
the dependence on the renormalisation and factorisation scales are
sizeable ($\sim 33\%$). Furthermore, the NLO corrections are rather large
($\sim 77\%$) indicating that predictions at NNLO QCD accuracy are
needed, although such computations are not within our reach with the
current technology. We note that the cross section ratio between
$t\bar{t}b\bar{b}$ and $t\bar{t}jj$ processes
($\sigma_{t\bar{t}b\bar{b}}/\sigma_{t\bar{t}jj}$) has also been
studied in Ref.~\cite{Bevilacqua:2014qfa}. Constructing such a ratio might
help reducing the theoretical uncertainties if the two processes are
correlated\footnote{Examples where this situation occurs are the recent
studies of the ratio of the  correlated processes
$t\bar{t}\gamma/t\bar{t}$~\cite{Bevilacqua:2018dny} and
$t\bar{t}W^+/t\bar{t}W^-$~\cite{Bevilacqua:2020srb}.}.  It was found
that $t\bar{t}b\bar{b}$ and $t\bar{t}jj$ are uncorrelated due to their 
different jet kinematics and consequently the scale uncertainty is not
significantly reduced when taking the ratio of the cross sections. To
understand the origin of these large NLO corrections and gain a
glimpse into the accuracy beyond NLO calculations, $t\bar{t}b\bar{b}$
production with an additional jet  ($pp\to t\bar{t}b\bar{b}j+X$) has
been studied in Ref.~\cite{Buccioni:2019plc}.

An accurate description of top-quark decays is essential to learn as
much information as possible about the top-quark properties from its
decay products.  The inclusion of top-quark decays has been first
performed through the matching of the NLO fixed-order
$t\bar{t}b\bar{b}$ calculation to parton shower programs (NLO+PS)
either in the 5-~\cite{Kardos:2013vxa,Garzelli:2014aba} or 4-flavour
scheme~\cite{Bevilacqua:2017cru, Jezo:2018yaf}\footnote{We note that
in Ref.~\cite{Cascioli:2013era} a NLO+PS computation for
$t\bar{t}b\bar{b}$ production has been performed without including
top-quark decays. The tool presented in that paper, however, supports
automated top-quark decays with spin correlations.}. In
Ref.~\cite{Jezo:2018yaf} even spin-correlated LO top-quark decays have
been included.  Although NLO+PS predictions allow a
resummation of large logarithms via the successive soft-collinear
emissions within the parton shower evolution, they account only for
double-resonant top-quark contributions, meaning that the single- and
non-resonant ones, which can be very important in some phase-space
regions, as well as their interference effects, are omitted.
Finally, for the simulations where the extra pair of $b$ jets is only
produced in the production stage of the top quarks (e.g.
$t\bar{t}b\bar{b}$+PS simulations), contributions from top-quark
decays with multi-bottom final states (i.e. $t \to W^+bb\bar{b}$ and
$\bar{t} \to W^- \bar{b}b\bar{b}$) are only described in PS
approximation. We also note that the
$t\bar{t}b\bar{b}$ background to $t\bar{t}H$ also involves
$t\bar{t}b\bar{b}g$ with $g\to b\bar{b}$, that can be consistently
included in NLO+PS simulations of $pp\to t\bar{t}b\bar{b}$ in the
4-flavour scheme.

The inclusive fiducial cross sections of $t\bar{t}b\bar{b}$ production
at the LHC and the cross section ratio
$\sigma_{t\bar{t}b\bar{b}}/\sigma_{t\bar{t}jj}$ have been measured in
the dilepton and lepton+jets decay channels at 8
TeV and 13 TeV by both ATLAS and CMS
experiments~\cite{CMS:2013vui,ATLAS:2018fwl,CMS:2017xnm,
CMS:2020grm}. Although the theoretical predictions, provided by the
NLO+PS simulations for inclusive fiducial cross sections and by the
fixed-order calculation of Ref.~\cite{Bevilacqua:2014qfa} for the
cross section ratio, are smaller than the measured ones, they are
still compatible with the measurements within the total uncertainties.

The state-of-the-art fixed-order computation for $pp\to
t\bar{t}b\bar{b}$ production in dilepton channel comprises NLO
QCD corrections at $\mathcal{O}(\alpha_s^5\alpha^4)$ with full
off-shell effects included.  In this calculation, the top quarks and
$W$ bosons are described by Breit-Wigner propagators and all the
double-, single- and non-resonant contributions (hereafter referred to
as DR, SR and NR) as well as the interference effects are consistently
incorporated at the matrix element level.  The full off-shell
calculation for $pp\to e^+ \nu_e \mu^- \bar{\nu}_\mu\,
b\bar{b}b\bar{b}+X$ has been carried out recently by two independent
groups~\cite{Denner:2020orv,Bevilacqua:2021cit}.  In
Ref.~\cite{Bevilacqua:2021cit}, the contribution of initial state
bottom quarks and PDF uncertainties have been additionally
investigated in detail.

In this paper, we examine the size of full off-shell effects in
$t\bar{t}b\bar{b}$ production by comparing different ways of
treating top-quark decays in
perturbative QCD. To achieve this we employ the NLO computation
performed in the narrow width approximation (NWA) with full spin
correlations included.  These are indeed the first predictions in the
NWA fully accurate at NLO in QCD for the $pp\to e^+ \nu_e \mu^- \bar{\nu}_\mu
b\bar{b}b\bar{b}+X$ process.  In the NWA, the top quarks and the $W$ bosons are
produced on-shell and decayed subsequently, which means that only the DR
contributions are taken into account.  We also provide predictions for the NWA
where the NLO QCD corrections are included in the top-quark production stage
but not in top-quark decays.  The NWA also helps us to understand the origin of
the four $b$ jets.  In this paper we work out a prescription, similar to
Refs.~\cite{Denner:2014wka,Denner:2020orv,
Bevilacqua:2020srb,Bevilacqua:2019quz}, to distinguish the following two types
of $b$ jets:
\begin{itemize}
  \item prompt $b$ jets which are originating from  $g\to b\bar{b}$
    splittings,
  \item $b$ jets from top-quark decays, i.e. $t\to W^+ b$ or $\bar{t}\to
W^-\bar{b}$.
\end{itemize}
 We will use our prescription to perform the labelling in the more realistic
full off-shell case and compare it to the NWA.  The prompt $b$-jet
identification is crucial because it enters the reconstruction of the Higgs
boson in $t\bar{t}H (H\to b\bar{b})$ measurements.  In that case an additional
condition must be satisfied, as shown for example in Ref.~\cite{ATLAS:2021qou}.
Namely, the reconstructed invariant mass of the Higgs boson should be in the
range of $M(bb) \in (100 - 140)$ GeV.  In general, our prescription for the
$b$-jet identification in $t\bar{t}b\bar{b}$ might help to improve the
understanding of the signal in $t\bar{t}H (H\to b\bar{b})$ measurements.

The paper is organised as follows. In Section~\ref{sec:setup}, the
theoretical setup for the LO and NLO QCD calculations is given.  We
then briefly review the NWA framework in Section~\ref{sec:nwa} and
describe our prescription to label the $b$ jets in
Section~\ref{sec:toprecon}.  In Section~\ref{sec:intxsec} we present
our findings for the integrated fiducial cross sections. This is
followed by the presentation of the results at a differential level in
Section~\ref{sec:diffxsec}, with particular emphasis on full off-shell
effects for a number of observables, as well as the results for
$b$-jet labelling. Finally, in Section~\ref{sec:summ} the results of
our paper are summarised.


\section{Description of the calculation and LHC setup}
\label{sec:setup}

    
We consider the  $pp\to e^+ \nu_e \mu^- \bar{\nu}_\mu b\bar{b}
b \bar{b} + X$ process at NLO in QCD for the LHC operating at a
center-of-mass
energy of $\sqrt{s}=13$ TeV.  In this paper we provide predictions for
this process based on the full off-shell calculation as well as on the NWA. Both
predictions have been obtained with the \textsc{Helac-Nlo} package
\cite{Bevilacqua:2011xh}, which is built around
\textsc{Helac-Phegas}~\cite{Cafarella:2007pc}.  In the
\textsc{Helac-Nlo} framework, the phase-space integration is performed
and optimised with \textsc{Parni}~\cite{vanHameren:2007pt} and
\textsc{Kaleu}~\cite{vanHameren:2010gg}.  The virtual
corrections are computed with
\textsc{Helac-1Loop}~\cite{vanHameren:2009dr}, which incorporates
\textsc{CutTools}~\cite{Ossola:2006us,Ossola:2007ax} and
\textsc{OneLOop} \cite{vanHameren:2010cp} as its cornerstones. The
\textsc{Helac-Dipoles} program~\cite{Czakon:2009ss} is employed to
evaluate the real corrections.  For the full off-shell calculation,
the real corrections are computed using both 
Catani-Seymour~\cite{Catani:1996vz,Catani:2002hc} and 
Nagy-Soper~\cite{Bevilacqua:2013iha} subtraction schemes. In the NWA
computation the Catani-Seymour scheme is used for the subtraction of
infrared divergencies in the production process, while in the case of
top-quark decays, the subtraction is performed according to the scheme
of Ref.~\cite{Campbell:2004ch} (see also \cite{Bevilacqua:2019quz} for
details). In computing real emission contributions, we employ a
restriction on the phase space of the dipoles, using the so-called
$\alpha_{\rm max}$ parameter, as originally proposed for the
Catani-Seymour scheme in Refs.~\cite{Nagy:1998bb,Nagy:2003tz} and for
the Nagy-Soper one in Ref.~\cite{Czakon:2015cla}.  The results must be
independent of the choice of the $\alpha_{\rm max}$ parameter, and
therefore, we choose two different values and check this independence.
To optimise the performance of \textsc{Helac-Nlo}, we use reweighting
techniques and MC sampling methods over the helicities and 
colours.  The results of our full off-shell calculation are stored in
the form of modified Les Houches Event Files~\cite{Alwall:2006yp} and
ROOT Ntuples~\cite{Antcheva:2009zz}. These event samples contain
matrix-element and PDF information which allows us to reweight 
to different scale settings and PDF sets.  Furthermore, working with
event samples enables us to analyse different sets of kinematical cuts,
as long as the phase-space region defined by the new set of cuts lies
within the original one. Finally, also new infrared-safe observables
can be produced without the need for additional time-consuming runs.

In this calculation we consider massless $b$ quarks and work in the 5
flavour scheme. Contributions from subprocesses with $b$ quarks in the
initial states are ignored given their numerical insignificance, as
shown in Ref.~\cite{Bevilacqua:2021cit}.  We keep the
Cabibbo-Kobayashi-Maskawa matrix diagonal and use the same SM
input parameters as  in Ref.~\cite{Bevilacqua:2021cit}
\begin{equation}
\begin{array}{lll}
  G_{F}=1.16638 \cdot 10^{-5} ~{\rm GeV}^{-2}\,,
  &\quad \quad \quad &   m_{t}=173 ~{\rm GeV} \,,
\vspace{0.2cm}\\
  m_{W}= 80.351972  ~{\rm GeV} \,, &
                     &\Gamma^{\rm NLO}_{W} =  2.0842989  ~{\rm GeV}\,,
\vspace{0.2cm}\\
  m_{Z}=91.153481   ~{\rm GeV} \,, &
                     &\Gamma^{\rm NLO}_{Z} = 2.4942664~{\rm GeV}\,.\\
\end{array}
\end{equation}
As leptonic $W$ boson decays are not subject to QCD corrections at the
one-loop level, we employ the total decay widths for $W$ and $Z$
bosons at NLO accuracy in our calculation, i.e. for LO and NLO matrix
elements. In our calculation all intermediate massive particles are
consistently treated in the complex-mass scheme
\cite{Denner:1999gp,Denner:2005fg}, where the widths of unstable
particles are absorbed into the imaginary part of the corresponding
mass parameters
\begin{equation}
 \mu_i^2 = m_i^2 - im_i\Gamma_i\;, \quad \text{for} \quad i \in \{W,Z,t\}\;.
\end{equation}
This choice implies a complex-valued weak mixing angle $\sin^2\theta_W = 1 -
\mu_W^2/\mu_Z^2$ and guarantees gauge invariance at NLO.
Complex masses are used through out in the propagators and in all couplings,
with the only exception of the electroweak coupling $\alpha$. The latter is
derived from the gauge-boson masses and the Fermi constant $G_F$ in the
$G_\mu$ scheme using the relation 
\begin{equation}
\alpha=\frac{\sqrt{2}}{\pi} \,G_{F} \,m_W^2 \left(
  1-\frac{m_W^2}{m_Z^2} \right)\,,
\end{equation}
namely real masses are employed therein.  Potential issues connected
to the usage of a complex value of the $\alpha$ coupling have been
discussed in the literature, see
e.g. Ref.~\cite{Frederix:2018nkq}. However, using complex masses
instead of real ones in the derived $\alpha$ does not impact our
predictions.  We have checked explicitly that using complex masses
instead of real ones in the derivation of $\alpha$ impacts our
predictions at the level of $0.2\%$, i.e. it is within the MC errors.
The top-quark width used for the off-shell calculation at LO,
$\Gamma^{\rm LO}_{t, {\rm off-shell}}$, is computed according to
Ref.~\cite{Jezabek:1988iv}.  The corresponding parameter for the NLO
calculation, $\Gamma^{\rm NLO}_{t, {\rm off-shell}}$, is computed by
applying the procedure of Ref.~\cite{Basso:2015gca} to the LO
width. The numerical values are as follows
\begin{equation}
  \Gamma_{t, {\rm off-shell}}^{\rm LO} =  1.443303  ~{\rm GeV}\,,
  \qquad \qquad \qquad
  \Gamma_{t, {\rm off-shell}}^{\rm NLO} =  1.3444367445   ~{\rm GeV}\,.
\end{equation}
A similar procedure applies in the NWA case, where in addition the
following limit is taken $\Gamma_W/m_W \to 0$ as shown in
Ref.~\cite{Denner:2012yc}. In this case we get
\begin{equation}
  \Gamma^{\rm LO}_{t,{\rm NWA}} = 1.466332  ~{\rm GeV}\,,
  \qquad \qquad \qquad
\Gamma^{\rm NLO}_{t,{\rm NWA}} =  1.365888 ~{\rm GeV}.
\end{equation}
We employ the following dynamical scale setting $\mu_R=\mu_F=\mu_0 =
H_T/3$ as our central scale, where $H_T$ is defined as

\begin{equation}
  H_T = \sum_{i=1}^{4} p_T(b_i) + p_T(e^+) + p_T(\mu^-) + p_T^{miss},
\end{equation}
and $p_T^{miss}$ is the  missing transverse momentum from the two
neutrinos. For the full off-shell calculation, the theoretical
uncertainty stemming from the scale dependence is obtained using the
standard $7$-point scale variation
\begin{equation}
  \biggl( \frac{\mu_R}{\mu_0}, \frac{\mu_F}{\mu_0} \biggr)
  = \Bigl \{ (2,1),(0.5,1),(1,2),(1,1),(1,0.5),(2,2),(0.5,0.5) \Bigr
  \} \,.
\end{equation}
For the NWA calculation, we only use the following $3$-point scale
variation
\begin{equation}
  \biggl( \frac{\mu_R}{\mu_0}, \frac{\mu_F}{\mu_0} \biggr)
  = \Bigl \{ (1,1),(2,2),(0.5,0.5) \Bigr \} \,.
\end{equation}
We note here that, as shown in Ref.~\cite{Bevilacqua:2021cit}, the
scale variation is driven mainly by the changes in $\mu_R$. Hence, the
uncertainties will not change between the $3$- and $7$-point scale
variations.  We would like to point out that we do not vary
$\alpha_s(m_t)$ entering $\Gamma_t^{\rm NLO}$ when computing the scale
dependence. We checked explicitly using the NWA that such effects are
small and increase the uncertainty estimates by at most $1\%$. We use
the LHAPDF6~\cite{Buckley:2014ana} interface and employ NNPDF3.1
\cite{NNPDF:2017mvq} as the default PDF set.  Specifically, we use the
following two PDF sets NNPDF31$\_$lo$\_$as$\_$0118 and
NNPDF31$\_$nlo$\_$as$\_$0118 for the LO and NLO calculation,
respectively. Both PDF sets are obtained with
$\alpha_s(m_Z)=0.118$. We apply the following cuts to the charged
leptons and $b$ jets:
\begin{equation}
  p_T(\ell)>20~\text{GeV}\,,\hspazio |y(\ell)|<2.5\,,\hspazio
  p_T(b)>25~\text{GeV}\,,\hspazio |y(b)|<2.5\,,
\end{equation}
where $\ell = e^+, \mu^-$.  All final-state partons with
pseudorapidity $|\eta|<5$ are recombined into jets via the anti-$k_T$
jet algorithm \cite{Cacciari:2008gp} with $R = 0.4$, where the
four-momentum recombination scheme is employed.  We require
exactly four  $b$ jets in the final state as well as two
charged leptons.  We put no restrictions on the extra light jet and
$p_T^{miss}$.


\section{The Narrow Width Approximation}
\label{sec:nwa}

\begin{figure}
\begin{center}
{\includegraphics[width=1\textwidth]{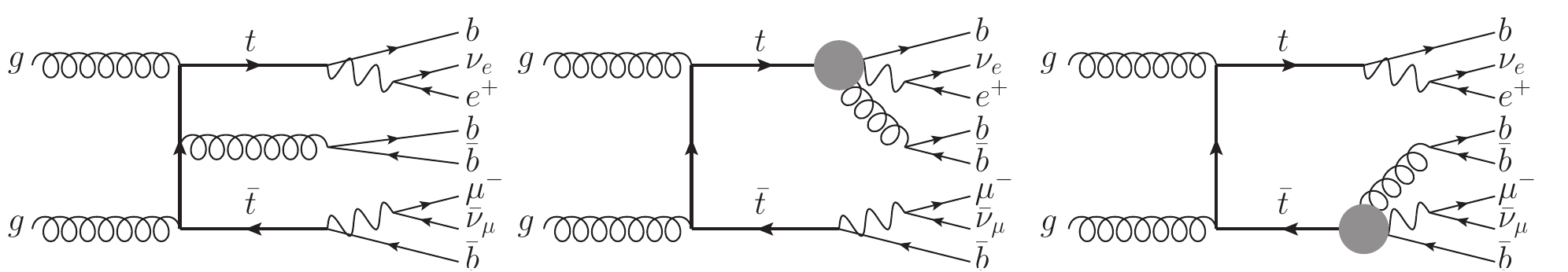}}
\caption{\it Representative tree-level Feynman diagrams for the  $pp\to
e^+ \nu_e  \, \mu^- \bar{\nu}_\mu \, b\bar{b} \,b\bar{b}$ process in
the NWA. The blob represents the $g\to b\bar{b}$ splitting in
top-quark decays. The latter can be emitted   either
from the top quark or the bottom quark.}
\label{feyndiag}
\end{center}
\end{figure}

In this Section we briefly review the NWA and highlight some of its features
that will be exploited throughout our study.  We start with the
propagator of a massive unstable particle appearing in the scattering
amplitude. The latter  is of the form
\begin{equation}
  \frac{1}{p^2-m^2+i m \Gamma},
\end{equation}
where $p$, $m$ and $\Gamma$ are the four-momentum, mass and decay
width of the particle, respectively.  When computing matrix elements,
we deal with the squared  propagator, which is given by the following
Breit-Wigner distribution
\begin{equation}
  \frac{1}{(p^2-m^2)^2+m^2\Gamma^2}.
\end{equation}
In the limit $\Gamma/m \to 0$, we obtain
\begin{equation}
  \label{BW-delta}
  \lim_{\Gamma/m \rightarrow 0} \frac{1}{(p^2-m^2)^2
    +m^2\Gamma^2} =
  \frac{\pi}{m\Gamma}\delta(p^2-m^2) \,.
\end{equation}
For the top quark and $W$ gauge boson, we have with our input parameters
\begin{equation}
\frac{\Gamma_t}{m_t} \approx 0.008, \spazio \frac{\Gamma_W}{m_W} \approx 0.026.
\end{equation}
The Dirac delta function in Eq.~\eqref{BW-delta} enforces the on-shell production
of both the top quarks and  $W$ gauge bosons.  This induces a factorization of the
cross section into a production and decay stage.  Therefore, we can write the
LO cross section in the NWA as follows 
\begin{align}
\label{ttbbNWALO}
d\sigma^{\rm LO}_{\rm NWA_{full}} = & \quad
 d\sigma^{\rm LO}_{t\bar{t}b\bar{b}}\frac{d\Gamma^{0}_{t\to W^+b}
  d\Gamma^{0}_{\bar{t}\to W^-\bar{b}}}{\bigl(\Gamma_{t,{\rm NWA}}^{\rm LO}\bigr)^2}
\nonumber \\
& + d\sigma^{\rm LO}_{t\bar{t}}\left(\frac{d\Gamma^{0}_{t\to W^+bb\bar{b}}
d\Gamma^{0}_{\bar{t}\to W^-\bar{b}}}{\bigl(\Gamma_{t,{\rm NWA}}^{\rm LO}\bigr)^2}
 +\frac{d\Gamma^{0}_{t\to W^+b}
d\Gamma^{0}_{\bar{t}\to W^-\bar{b}b\bar{b}}}{\bigl(\Gamma_{t,{\rm
               NWA}}^{\rm LO}\bigr)^2}
               \right),
\end{align}
where $d\sigma^{\rm LO}_{t\bar{t}b\bar{b}}$ ($d\sigma^{\rm
LO}_{t\bar{t}}$) is the cross section of on-shell $t\bar{t}b\bar{b}$
($t\bar{t}$) production at LO and $d\Gamma^{0}_{{t}\to W^+b}$
($d\Gamma^{0}_{\bar{t}\to W^-\bar{b}}$, $d\Gamma^{0}_{{t}\to
W^+b{b}\bar{b}}$, $d\Gamma^{0}_{\bar{t}\to W^-\bar{b}b\bar{b}}$) is
the differential decay rate of $t\to W^+b$ (${\bar{t}\to W^-\bar{b}}$,
$t\to W^+bb\bar{b}$, ${\bar{t}\to W^-\bar{b}b\bar{b}}$) at LO.  We
stress here that the full spin correlations are incorporated.  The
three terms in Eq.~\eqref{ttbbNWALO} represent the three corresponding
subprocesses depicted in Figure~\ref{feyndiag}.
%
%
At NLO, we can write the cross section in the NWA as follows
\begin{eqnarray}
\label{ttbbNWANLO}
  d\sigma^{\rm NLO}_{\rm NWA_{full}} &=& 
  d\sigma^{\rm NLO}_{t\bar{t}b\bar{b}}\frac{d\Gamma^{0}_{t\to W^+b}
  d\Gamma^{0}_{\bar{t}\to W^-\bar{b}}}{\bigl(\Gamma_{t,{\rm NWA}}^{\rm NLO}\bigr)^2}
        \nonumber\\
&&+\,d\sigma^{\rm NLO}_{t\bar{t}}\left(\frac{d\Gamma^{0}_{t\to   W^+bb\bar{b}}
d\Gamma^{0}_{\bar{t}\to W^-\bar{b}}}{\bigl(\Gamma_{t,{\rm NWA}}^{\rm NLO}\bigr)^2}
   +\frac{d\Gamma^{0}_{t\to W^+b}
d\Gamma^{0}_{\bar{t}\to W^-\bar{b}b\bar{b}}}{\bigl(\Gamma_{t,{\rm
   NWA}}^{\rm NLO}\bigr)^2} \right)  \nonumber\\
&&+\,d\sigma^{\rm LO}_{t\bar{t}b\bar{b}}\left(\frac{d\Gamma^{1}_{t\to   W^+b}
d\Gamma^{\rm 0}_{\bar{t}\to W^-\bar{b}}}{\bigl(\Gamma_{t,{\rm
   NWA}}^{\rm NLO}\bigr)^2}
   +\frac{d\Gamma^{0}_{t\to W^+b}
 d\Gamma^{1}_{\bar{t}\to W^-\bar{b}}}{\bigl(\Gamma_{t,{\rm NWA}}^{\rm NLO}\bigr)^2}\right)
   \nonumber\\
&&+\,d\sigma^{\rm LO}_{t\bar{t}}\left( \quad \frac{d\Gamma^{1}_{t\to   W^+bb\bar{b}}
 d\Gamma^{0}_{\bar{t}\to W^-\bar{b}}}{\bigl(\Gamma_{t,{\rm NWA}}^{\rm   NLO}\bigr)^2}
   +\frac{d\Gamma^{1}_{t\to W^+b}
d\Gamma^{0}_{\bar{t}\to W^-\bar{b}b\bar{b}}}{\bigl(\Gamma_{t,{\rm   NWA}}^{\rm NLO}\bigr)^2}
\right. \nonumber \\
        & & \qquad\qquad\;\; + \left. \frac{d\Gamma^{0}_{t\to  W^+bb\bar{b}}
d\Gamma^{1}_{\bar{t}\to W^-\bar{b}}}{\bigl(\Gamma_{t,{\rm NWA}}^{\rm NLO}\bigr)^2}
            +\frac{d\Gamma^{0}_{t\to W^+b}
 d\Gamma^{1}_{\bar{t}\to W^-\bar{b}b\bar{b}}}{\bigl(\Gamma_{t,{\rm  NWA}}^{\rm NLO}\bigr)^2}
           \right),
\end{eqnarray}
where $d\sigma^{\rm NLO}_{t\bar{t}b\bar{b}}$ ($d\sigma^{\rm
NLO}_{t\bar{t}}$) is the cross section of on-shell $t\bar{t}b\bar{b}$
($t\bar{t}$) production at NLO and $d\Gamma^{1}_{t\to W^+b}$
($d\Gamma^{1}_{\bar{t}\to W^-\bar{b}}$, $d\Gamma^{1}_{{t}\to
W^+b{b}\bar{b}}$, $d\Gamma^{1}_{\bar{t}\to W^-\bar{b}b\bar{b}}$) is
the NLO correction to the differential decay rate of $t\to W^+b$
(${\bar{t}\to W^-\bar{b}}$, $t\to W^+bb\bar{b}$, ${\bar{t}\to
W^-\bar{b}b\bar{b}}$).  From Eq.~\eqref{ttbbNWANLO} we can see that
the NLO corrections can occur either in the production stage of the
top quarks or in their decays.  Moreover, similar to the LO
calculation, prompt $b$ jets can be produced in both stages. It is
worth to point out that the various terms in Eq.~(\ref{ttbbNWANLO})
are separately infrared finite and do not interfere with each other.

In this work we consider predictions based on the NWA using various levels
of accuracy, as described below:
\begin{itemize}
  \item $\rm NWA_{full}$: all contributions in Eq.~\eqref{ttbbNWANLO} are taken
into account.
  \item $\rm NWA_{LOdec}$: QCD corrections to  top-quark decays are
neglected.  This prediction is obtained by considering only the first and second
lines in Eq.~\eqref{ttbbNWANLO} and using $\Gamma^{\rm LO}_{t,{\rm NWA}}$ in
the calculation.
  \item $\rm NWA_{prod}$: contributions where the prompt $b$ jets originating
from top-quark decays are neglected.  To obtain this prediction we consider
only the first and third lines in Eq.~\eqref{ttbbNWANLO}.
  \item $\rm NWA_{LOdec,prod}$: both the QCD corrections to top-quark decays and
the contributions of prompt $b$ jets from top-quark decays are neglected.  The
contribution to this result comes from the first line of Eq.~\eqref{ttbbNWANLO}
only, using $\Gamma^{\rm LO}_{t,{\rm NWA}}$ in the calculation.
\end{itemize}
We note that the presence in Eq.~\eqref{ttbbNWANLO} of the NLO
top-quark width, which is a function of the strong coupling constant,
spoils the rigorous expansion of the cross section in powers of
$\alpha_s$.  As pointed out, for example in
Ref.~\cite{Melnikov:2009dn,Melnikov:2011ta,Melnikov:2011qx,Behring:2019iiv,
Czakon:2020qbd}, a consistent expansion of
$d\sigma^{\rm NLO}_{\rm NWA_{full}}$ in the strong coupling constant
should be performed. This expansion leads to the following definition
of the so-called expanded NWA  cross section (further denoted by $\rm
NWA_{exp}$)
\begin{equation}
 \label{eq:expandedNWA}
 d\sigma^{\rm NLO}_{\rm {NWA_{exp}}}
 = d\sigma^{\rm NLO}_{\rm {NWA_{full}}}
 \left( \frac{\Gamma^{\rm NLO}_{t,{\rm NWA}}}{\Gamma^{\rm LO}_{t,{\rm
         NWA}}}\right)^2
 -d\sigma^{\rm LO}_{\rm {NWA_{full}}}\frac{2(\Gamma^{\rm NLO}_{t,{\rm
       NWA}}
   -\Gamma^{\rm LO}_{t,{\rm NWA}})}{\Gamma^{\rm LO}_{t,{\rm NWA}}},
\end{equation}
where $d\sigma^{\rm LO}_{\rm {NWA_{full}}}$ should be computed with
$\Gamma_{t,{\rm NWA}}^{\rm LO}$ but employing NLO PDF sets.

Several comments are in order at this stage. The complex-mass
scheme used in the full off-shell calculation is by construction gauge
invariant and unitary to all orders in perturbation theory. Unitarity
is violated only at fixed order due to the truncation of perturbation
theory. However, the size of the violation is always a higher order
effect. Due to gauge invariance, one does not expect unnecessary
enhancements of unitarity violation
\cite{Denner:2014zga,Denner:2006ic}. Also in the NWA, for the dynamical
scale setting ($\mu_{R} = \mu_{F} = \mu_0 =H_T/3$) there is a
unitarity violation, see e.g. Ref. \cite{Czakon:2020qbd}.  But, again, it is
of higher orders. Generally, all three approaches (full off-shell
calculation, ${\rm NWA}_{\rm full}$ and ${\rm NWA}_{\rm exp}$) are a proper
description of the physics of the process under consideration. They differ,
of course, in the effects they contain, the full off-shell calculation
being the complete one compared to the NWA results. What remains to be
discussed is which version of the NWA should be compared to the full
off-shell calculation in order to quantify the size of non-factorisable
corrections.  In NLO QCD calculations in the NWA there are various
possibilities for treating $\Gamma_t$. For example, one can set it to the
numerical value corresponding to the perturbative order of the full
calculation, see {\it e.g.} Ref.  \cite{Melnikov:2011qx}.  Alternatively, one
can formally expand it according to Eq.(\ref{eq:expandedNWA}). This expansion
has the effect of  removing systematically contributions corresponding to QCD
corrections applied in both production and decays simultaneously. The latter
represent formally higher-order contributions relative to the NLO QCD
perturbative accuracy.  We note that the size of these effects can be enhanced
for processes with large NLO QCD corrections in the production part.  For the
$t\bar{t}b\bar{b}$ process under consideration, assuming that contributions
from $g \to b\bar{b}$ splittings in top-quark decays are sufficiently small to
be negligible, these effects are expected to be of the order of twice the
product of the NLO QCD corrections to production and to the decay width of the
top quark. Being part of the higher-order corrections to the NLO cross section,
these effects should be smaller than the estimated scale uncertainties at the
perturbative order considered in the calculation. For the benchmark NWA results
shown in this study we will consider the results based on unexpanded top-quark
width.  The reason for not using this expansion for the results in the NWA
should be rather clear as such a procedure can not be directly applied to the
full off-shell calculations. Because the main purpose of the paper is a
consistent comparison between the NWA and the full off-shell results, such
approach seems to be more appropriate, see also Refs.
\cite{SM:2012sed,Bevilacqua:2020srb,Bevilacqua:2019quz,
Bevilacqua:2020pzy,Hermann:2021xvs,Bevilacqua:2021tzp, Stremmer:2021bnk}.
However, when studying the integrated fiducial cross sections, we will also
report our findings based on the expanded approach for comparison.  Such a
comparison will be also helpful to assess the effective impact of QCD
corrections to top-quark decays in our analysis.

Finally, we would like to point out that the size of the top quark
non-factorisable corrections can be estimated directly from the full
off-shell result. This can be done by rescaling the following
couplings $t-W-b$ and $W-\ell-\nu_\ell$ by several large factors, as
described in Ref. \cite{Bevilacqua:2010qb,Denner:2012yc}. This
approach should mimic the $\Gamma_t \to 0$ limit when the scattering
cross section factorises into on-shell production and decay. We used
both approaches, the coupling rescaling and ${\rm NWA}_{\rm full}$, for
the simpler process, $pp \to e^+ \nu_e \, \mu^-\bar{\nu}_\mu\,
b\bar{b}\, \gamma +X$, and have shown that both methods give very
similar results, see e.g Ref.  \cite{Bevilacqua:2019quz}. As the
coupling rescaling requires the rerunning of the full off-shell
calculation, we of course can not afford to do this for the more
complicated $pp \to e^+ \nu_e\, \mu^-\bar{\nu}_\mu \, b\bar{b}\,
b\bar{b} +X$ process.

    
\section{Top-quark reconstruction and $b$-jet identification}
\label{sec:toprecon}


Experimental measurements of Higgs boson properties in the $pp\to
t\bar{t}H$ process with $H \to b\bar{b}$ require the identification of
the Higgs boson by combining the $b\bar{b}$ pair originating from its
decay.  From the viewpoint of realistic analysis, the final state
$t\bar{t}H\to W^+W^-\,b\bar{b}\,b\bar{b}$, where $W^+W^-$ can decay
fully leptonically $(\ell \nu_\ell \,\ell \nu_\ell)$,
semi-leptonically $(qq\,\ell \nu_\ell)$ or fully hadronically
$(qq\,qq)$, consists of multiple $b$ jets originating either from
top-quark decays or the Higgs boson. The $b$ jets which are not
associated with top-quarks decays are labelled prompt $b$
jets. However, the proper identification of prompt $b$ jets is not
free of ambiguities and can lead to substantial smearing of what would
be a sharp Higgs resonance peak in the distribution of the invariant
mass of the $b\bar{b}$ system. In this Section we provide a
prescription for labelling prompt $b$ jets in the $pp\to e^+ \nu_e \,
\mu^- \bar{\nu}_\mu \,b\bar{b}\,b\bar{b}+X$ process. Even though we
focus on the QCD $t\bar{t}b\bar{b}$ background, our conclusions can be
of interest also for the case of the $pp \to t\bar{t}H(H\to b\bar{b})$
signal.


\subsection{Full off-shell case}


To perform such a labelling, in the full off-shell case we attempt to
reconstruct the top quarks.  The method was introduced in
Ref.~\cite{Kauer:2001sp} and further discussed in e.g.
Refs.~\cite{Bevilacqua:2020srb,Bevilacqua:2019quz}.  In this paper, we
generalize this method in two aspects. In
Refs.~\cite{Bevilacqua:2020srb,Bevilacqua:2019quz}, the top-quark
reconstruction is based on the assumption that the charge of $b$ and $\bar{b}$
jets can be distinguished.  In our labelling procedure, first, we are going to
make use of the $b$-jet charge information, however, we are also going to
generalize this method to the case where $b$ and $\bar{b}$ cannot be
distinguished.  In Refs.~\cite{Bevilacqua:2020srb,Bevilacqua:2019quz} the top
quark is reconstructed by minimizing the function
${\cal Q}=|M(t)-m_t|+|M(\bar{t})-m_t|$. In this paper, we define the function
${\cal Q}$ as
\begin{equation} \label{Qvariable}
  {\cal Q}=|M(t)-m_t|\times|M(\bar{t})-m_t|\times|M^{\rm  prompt}(bb)|,
\end{equation}
where $M(t)$ and $M (\bar{t})$ are the invariant mass of the candidate
$t$ and $\bar{t}$, $M^{\rm prompt}(bb)$ is the invariant mass of the
candidate pair of prompt $b$ jets and $m_t = 173$ GeV.  Compared to
${\cal Q}=|M(t)-m_t|+|M(\bar{t})-m_t|$, the new definition adapts
better to the actual resonant structure of the matrix elements for the
process at hand.  For each event, we evaluate this function for all
possible resonant histories, namely all possible ways to reconstruct
top-quark momenta based on the particle content of the final
state. The most likely resonant history, that we take as reference to
reconstruct top quarks and thus to label prompt $b$ jets, is the one
which minimises the function ${\cal Q}$. We point out that a similar
prescription can be used for the identification of prompt $b$ jets in
the $pp \to t\bar{t} H(H \to b\bar{b})$ process as well. To this end
it is sufficient to replace $M^{\rm prompt}(bb)$ in
Eq.~\eqref{Qvariable} with $M^{\rm prompt}(bb) - m_H$, where $m_H$ is
the Higgs boson's nominal mass of $125$ GeV.


\subsubsection*{Charge-aware labelling}\label{caa}


Exploiting the information of the $b$-jet charge, $8$ resonant
histories are possible at LO. Denoting the final state as $W^+W^- b_1
\bar{b}_1 b_2 \bar{b}_2$ for ease of notation, the histories read as follows:
\begin{enumerate}
\item $t\rightarrow W^+ b_i,
  ~~~~~~~~~~\bar{t}\rightarrow W^- \bar{b}_j,
  ~~~~~~~~~~ i,j=1,2$;
\item $t\rightarrow W^+ b_1 b_2 \bar{b}_1,
  ~~~~\,\bar{t}\rightarrow W^- \bar{b}_2,
  ~~~~~~~~~~$and $1\leftrightarrow 2$;
\item $t\rightarrow W^+ b_1,
  ~~~~~~~~~~\bar{t}\rightarrow W^- \bar{b}_1 b_2 \bar{b}_2,
  ~~~~\,$and $1\leftrightarrow 2$.
\end{enumerate}
Here the subscripts are just used to distinguish the momenta of
identical particles and do not refer to any particular ordering. We
assume perfect reconstruction of the $W$ gauge bosons, i.e. $W^+ \to
e^+ \nu_e$ and $W^- \to \mu^- \bar\nu_\mu$. In the list above, the
resonant histories appearing in items $2$ and $3$ involve gluon
emission and a subsequent splitting into a bottom-quark pair in the
decay stage of the top quark and antitop quark, respectively. In this
case multiple $b$ jets enter the top-quark reconstruction, and the
definition of prompt $b$ jets as jets which are not originated from
top-quark decays clearly does not apply. Nevertheless, one can still
adopt a prescription to label the various $b$ jets to be prompt or
non-prompt.  Let us consider, for instance, $t\to W^+ b_1 b_2
\bar{b}_1$.  We can immediately identify $\bar{b}_1$ as a prompt $b$ jet
because it can only originate from a $g \to b\bar{b}$ splitting.
However in order to label $b_1$ and $b_2$, additional criteria are
needed. We propose two discriminators to achieve this goal:
\begin{enumerate}
  \item $p_T$ discriminator: the top-quark decay products are usually
    harder in $p_T$  than those of a $g\to b\bar{b}$ splitting.  Thus,
    we associate the  $b$ jet with the highest  $p_T$ to the top-quark
    decay and the other one to the prompt $b\bar{b}$ pair.
  \item  $\Delta R$  discriminator: the $b\bar{b}$ pair from the gluon splitting is
expected to have the smallest angular separation.  Thus, the $b$ jet that
minimises $\Delta R(b_i \bar{b}_1)$ $(i=1,2)$ will be labelled as the prompt
$b$ jet and the remaining one will be associated to the top-quark decay.
\end{enumerate}
At NLO the light jet from real radiation enters the definition of the
possible resonant configurations, and $16$ additional histories have
to be taken into account. We note that, to mimic what is
done on the experimental side, the light jet is included in the histories only
when it is resolved by the jet algorithm and satisfies the following
conditions: $p_T(j)>25$~GeV and $|y(j)|<2.5$.


\subsubsection*{Charge-blind labelling}\label{cba}


If we ignore the information about the charge of the $b$ jets,  we have the
following 20 histories at LO
\begin{enumerate}
\item $t\rightarrow W^+ b_i,~~~~~~~~~~
  \bar{t}\rightarrow W^- b_j,~~~~~~~~~~ i\neq j=1,2,3,4$;
\item $t\rightarrow W^+ b_1 b_2 b_3,~~~~\,\bar{t}\rightarrow
  W^- b_4,~~~~~~~~~~$and permutations;
\item $t\rightarrow W^+ b_1,~~~~~~~~~~\bar{t}\rightarrow
  W^- b_2 b_3 b_4,~~~~\,$and permutations,
\end{enumerate}
where we use the symbol $b$ for all the $b$ and $\bar{b}$ jets.  Again, the
subscripts do not refer to any ordering, but are introduced  to
distinguish particles. In the list above, the usage of  either the $p_T$
or $\Delta R$ discriminators is relevant for the histories reported in the last
two items. At NLO, there are $40$ additional histories due to real radiation.


\subsection{NWA case}


The NWA allows us to assess the effectiveness of our method for labelling the
$b$ jets. Indeed we can exploit both the facts that the top quarks are always
produced on-shell and that the cross section is factorized into
production and decays.  As a consequence of these properties, as shown in
Section~\ref{sec:nwa}, at LO we have three separate contributions to the 
cross section
\begin{enumerate}
\item $pp \rightarrow t\bar{t}b\bar{b}
  \rightarrow (e^+ \nu_e \, b) \; (\mu^- \bar{\nu}_{\mu} \, \bar{b}) \; b \bar{b}$
\item $pp \rightarrow t\bar{t}
  \rightarrow (e^+ \nu_e \, b b \bar{b}) \; (\mu^- \bar{\nu}_{\mu} \, \bar{b})$
\item $pp \rightarrow t\bar{t}
  \rightarrow (e^+ \nu_e \, b ) \; (\mu^- \bar{\nu}_{\mu} \, b \bar{b} \bar{b})$
\end{enumerate}
Therefore, we have full knowledge of the resonant history that is generated for
every phase-space point.  The only ambiguity left concerns contributions 2 and
3, where either the $p_T$ or $\Delta R$ discriminator is additionally needed to
label the $b$ jets. 
%
\begin{figure}[t!]
\centering
 \includegraphics[width=0.7\textwidth]{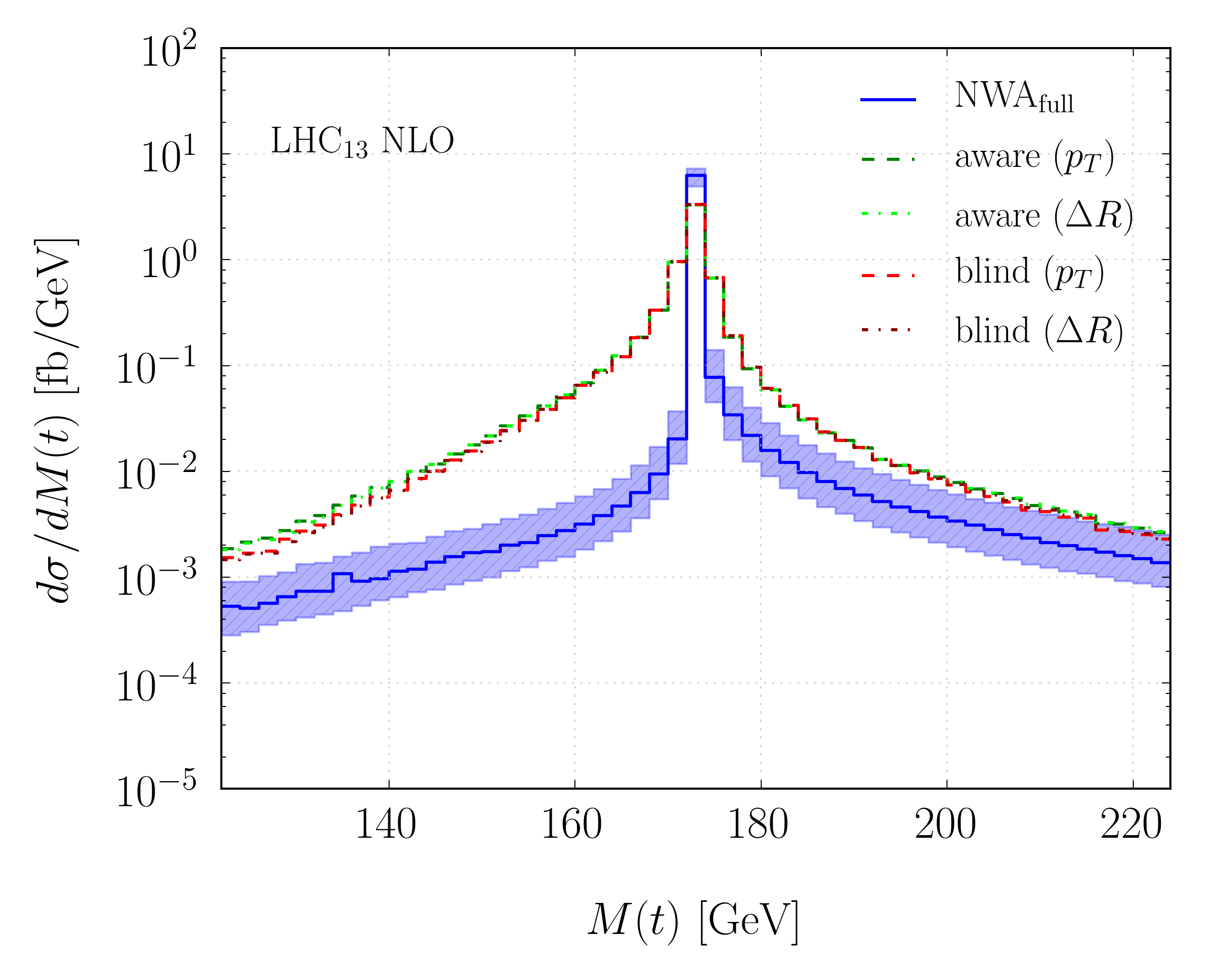}
\caption{\textit{Reconstructed invariant mass of the top quark at NLO in QCD
for $pp\to  e^+ \nu_e  \mu^- \bar{\nu}_\mu b\bar{b}b\bar{b}+X$ at the LHC with
$\sqrt{s}=13$ TeV. The NLO NNPDF3.1 PDF set and the dynamical scale, $\mu_R =
\mu_F = \mu_0=H_T/3$, are employed. We compare the full off-shell predictions,
obtained using the charge-aware and charge-blind labelling in combination with
different choices for the discriminator ($p_T$ and $\Delta R$), to the $\rm
NWA_{full}$ prediction.}}
\label{mt-nlo}
\end{figure}

The labelling performed on predictions in the NWA is used as a
reference for the more realistic full off-shell case.  In the NWA,
both the LO $M(t)$ and $M(\bar{t})$ distributions are simply given by
Dirac delta distributions, however, at NLO they are smeared by extra
radiation.  To validate the top-quark reconstruction we present in
Figure~\ref{mt-nlo} the $M(t)$ distribution at NLO in QCD in the
vicinity of the top-quark resonance. Similar results are obtained for
$M(\bar{t})$.  The $M(t)$ distribution strongly peaks around the
top-quark mass used in our calculation.  All predictions develop a
radiative tail due to final-state gluon radiation that is not
recombined with the top-quark decay products.  This effect is enhanced
in the case of the full off-shell calculation where also finite
top-quark width corrections as well as SR and NR top-quark
contributions and  interference effects play a crucial role.

    
\section{Integrated fiducial cross sections}
    \label{sec:intxsec}


In this Section we present theoretical predictions for integrated
fiducial cross sections for the $pp\rightarrow e^+ \nu_e \mu^-
\bar{\nu}_\mu b \bar{b} b \bar{b} + X$ process.  Comparing the NWA
predictions to the full off-shell results of
Ref.~\cite{Bevilacqua:2021cit} allows us to quantify the size of the
full off-shell effects. Moreover, armed with our top-quark
reconstruction method, we can approximately disentangle the double-,
single- and non-resonant contributions in the full off-shell
calculation by defining the corresponding phase-space regions, see
e.g.
Refs.~\cite{Kauer:2001sp,Liebler:2015ipp,Baskakov:2018huw,Bevilacqua:2019quz}.
We use the charge-aware scheme in the top-quark reconstruction and
define the DR region via the following conditions:
\begin{equation}\label{DR}
  |M(t)-m_t|<n\Gamma_t \hspazio \text{and} \hspazio
  |M(\bar{t})-m_t|<n\Gamma_t\,,  
\end{equation}
where we use $n=15$. Moreover,  $M(t)$, $M(\bar{t})$ are the same
quantities as those given in Eq.~\eqref{Qvariable}. There are two SR
phase-space regions, that are given by
\begin{equation}
  |M(t)-m_t|<n\Gamma_t \hspazio \text{and} \hspazio
  |M(\bar{t})-m_t|>n\Gamma_t\,,
\end{equation}
or
\begin{equation}
  |M(t)-m_t|>n\Gamma_t \hspazio \text{and} \hspazio
  |M(\bar{t})-m_t|<n\Gamma_t \,.
\end{equation}
Finally, the NR region is chosen according to
%
\begin{table}[t]
\centering
\begin{tabular}{ccccc}
  \hline
  Decay treatment &$\sigma^{\rm NLO}_i$ [fb] & $+\delta_{\rm scale}$ [fb]
  & $-\delta_{\rm scale}$ [fb]
  & $\sigma^{\rm NLO}_{i}/\sigma^{\rm NLO}_{\rm {NWA_{full}}}-1$
  \\[0.1cm]
  \hline
 Off-shell & $13.22~(2)$ & $+2.65~(20\%)$ & $-2.96~(22\%)$ & $+0.5\%$ \\
 DR        & $12.08~(2)$ & $-$             & $-$             & $-$ \\
SR & $1.112~(5)$ & $-$ & $-$ & $-$ \\
NR & $0.0249~(4)$ & $-$ & $-$ & $-$ \\
  $\rm NWA_{full}$ & $13.16~(1)$ & $+2.61~(20\%)$
  & $-2.93~(22\%)$ & $-$ \\
  $\rm NWA_{ LOdec}$ & $13.22~(1)$ & $+3.77~(29\%)$
  & $-3.31~(25\%)$ & $+0.5\%$ \\
  $\rm NWA_{exp}$ & $12.38~(1)$ & $+2.91~(24\%)$
  & $-2.89~(23\%)$ & $-5.9\%$ \\
  $\rm NWA_{ prod}$ & $13.01~(1)$ & $+2.58~(20\%)$
  & $-2.89~(22\%)$ & $-1.1\%$ \\
  $\rm NWA_{exp, prod}$  & $12.25~(1)$ & $+2.87~(23\%)$ 
  & $-2.86~(23\%)$ & $-6.9\%$ \\
  $\rm NWA_{ LOdec, prod}$ & $13.11~(1)$ & $+3.74~(29\%)$
  & $-3.28~(25\%)$ & $-0.4\%$ \\
  \hline
\end{tabular}
\caption{\it Integrated fiducial cross sections at NLO in QCD for
$pp\rightarrow e^+ \nu_e \mu^- \bar{\nu}_\mu b \bar{b} b \bar{b} + X$ at the
LHC with $\sqrt{s}=13$ TeV.  The NLO NNPDF3.1 PDF set and the dynamical scale,
$\mu_R = \mu_F = \mu_0=H_T/3$, are employed. The full off-shell prediction and its
DR, SR and NR contributions are presented along with various NWA
predictions. Also given are the theoretical uncertainties coming from scale variation
$(\delta_{\rm scale})$ and the relative differences to the $\rm
NWA_{full}$ prediction.}
\label{intxs}
\end{table}
%
\begin{equation}\label{NR}
  |M(t)-m_t|>n\Gamma_t \hspazio \text{and} \hspazio
  |M(\bar{t})-m_t|>n\Gamma_t \,.
\end{equation}
The integrated fiducial cross sections at NLO in QCD are presented in
Table~\ref{intxs}. Specifically, the full off-shell prediction and results
based on the NWA at various levels of accuracy (defined in
Section~\ref{sec:nwa}) are presented.  Comparing the full off-shell cross
section with the ${\rm NWA_{full}}$ result, we found that the full off-shell
effects are negligible for this process, i.e. they are below $0.5\%$.  The
theoretical uncertainties due to scale variation are of the order of $22\%$ and
are the same for both the full off-shell and the $\rm NWA_{full}$ calculations.
Comparing the results dubbed ${\rm NWA_{exp, prod}}$ and ${\rm NWA_{LOdec,
prod}}$ gives us an opportunity to assess the genuine impact of NLO QCD
corrections to top-quark decays in the $t\bar{t}b\bar{b}$ production picture.
These corrections are  negative and at the level of $6.6\%$.  On the other
hand, comparing  ${\rm NWA_{prod}}$ with ${\rm NWA_{LOdec, prod}}$ we observed
that the two results are fairly close to each other. We remark that this
agreement is accidental. Indeed the ${\rm NWA_{prod}}$ cross section includes
the formally suppressed higher-order contributions that have been discussed in
section \ref{sec:nwa}, which appear to be of similar size of the QCD correction
to decays. We want to emphasise that the latter conclusion is, to some extent,
dependent on the process under consideration and on the kinematical cuts
adopted for the analysis. In addition, the theoretical uncertainty increases up
to about $29\%$ when the NLO QCD corrections to top-quark decays are omitted.
By comparing the $\rm NWA_{prod}$ to the $\rm NWA_{full}$, we found that the $t
\to W^+ b\bar{b}b$ and $\bar{t} \to W^- \bar{b}b\bar{b}$ contributions are at
the $1.1\%$ level.  Finally, we observed that the $\rm NWA_{exp}$ cross section
is reduced by $5.9\%$ compared to the $\rm NWA_{full}$ result. Once more this
effect is ascribed to the formal higher-order contributions which are present
in $\rm NWA_{full}$. 
This leads to a difference of about
$6.4\%$ between the full off-shell case and the $\rm NWA_{exp}$ one.
The scale uncertainty for the $\rm NWA_{exp}$ is at the level of
$24\%$ which is still comparable to the full off-shell case.  We
conclude that the various NWA predictions, as well as the full
off-shell result, are in good agreement within the estimated
theoretical uncertainties. Also reported in Table~\ref{intxs} are the
contributions to the full off-shell cross section associated with the
DR, SR, NR regions. The DR region dominates with a $91.4\%$
contribution. It is followed by the SR region with $8.4\%$ and finally
the NR one at the level of $0.2\%$. Let us stress here, however, that
the relative size of the various contributions depends crucially on
the value of the $n$ parameter used in the definition of those
phase-space regions.
To conclude this part we note that, at the integrated fiducial level, all the
provided  theoretical predictions  for   $t\bar{t}b\bar{b}$ production in
dilepton channel give similar results.  The differences between  them are at
the level of a few percent at most, thus,  well within the estimated
theoretical uncertainties for this process, which are of the order of $22\%$.
Therefore, all of the predictions discussed in this section can be considered
equally valid at the integrated cross-section level.

    
\section{Differential fiducial cross sections}
    \label{sec:diffxsec}

    
\begin{figure}[t!]
\centering
\includegraphics[width=0.49\textwidth]{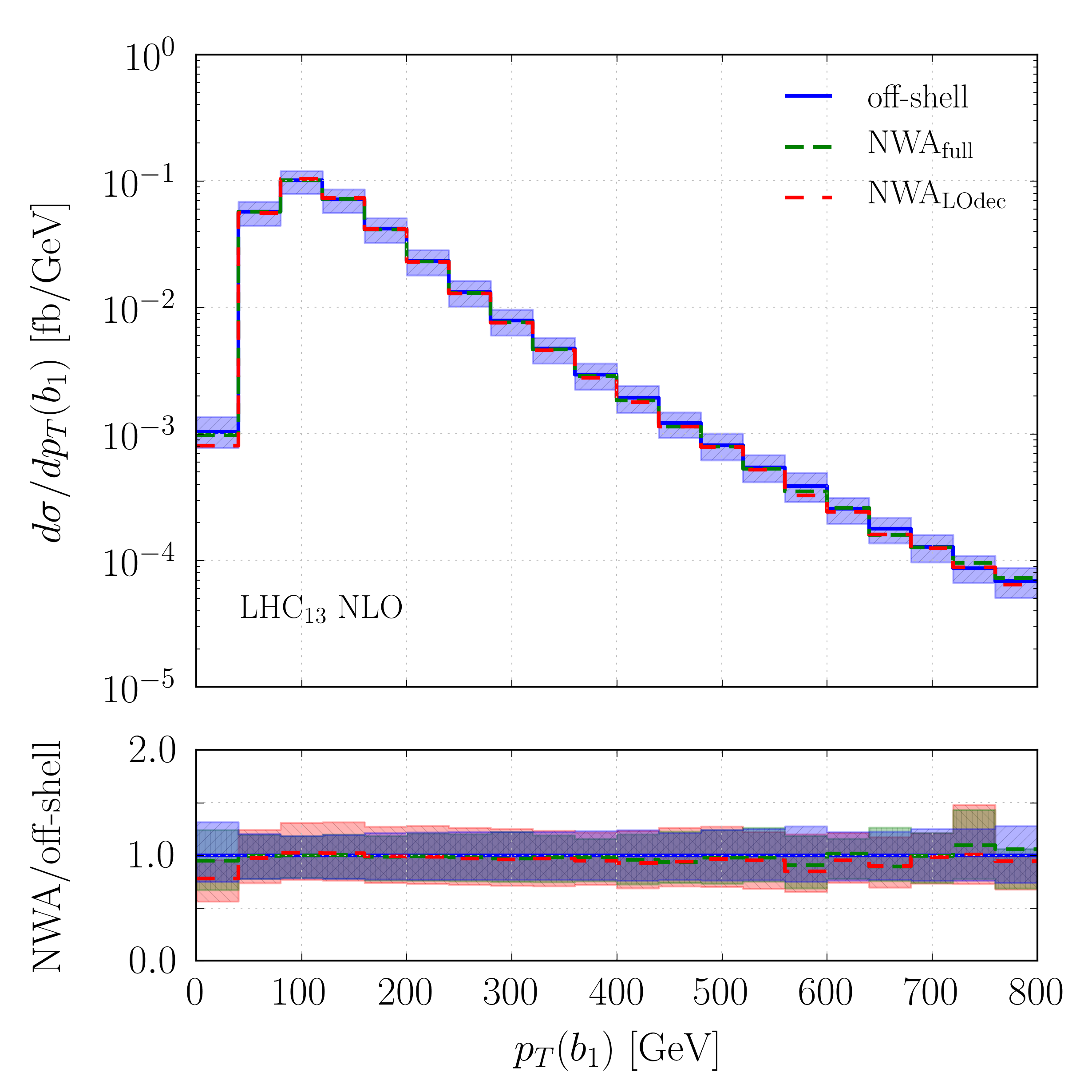}
\includegraphics[width=0.49\textwidth]{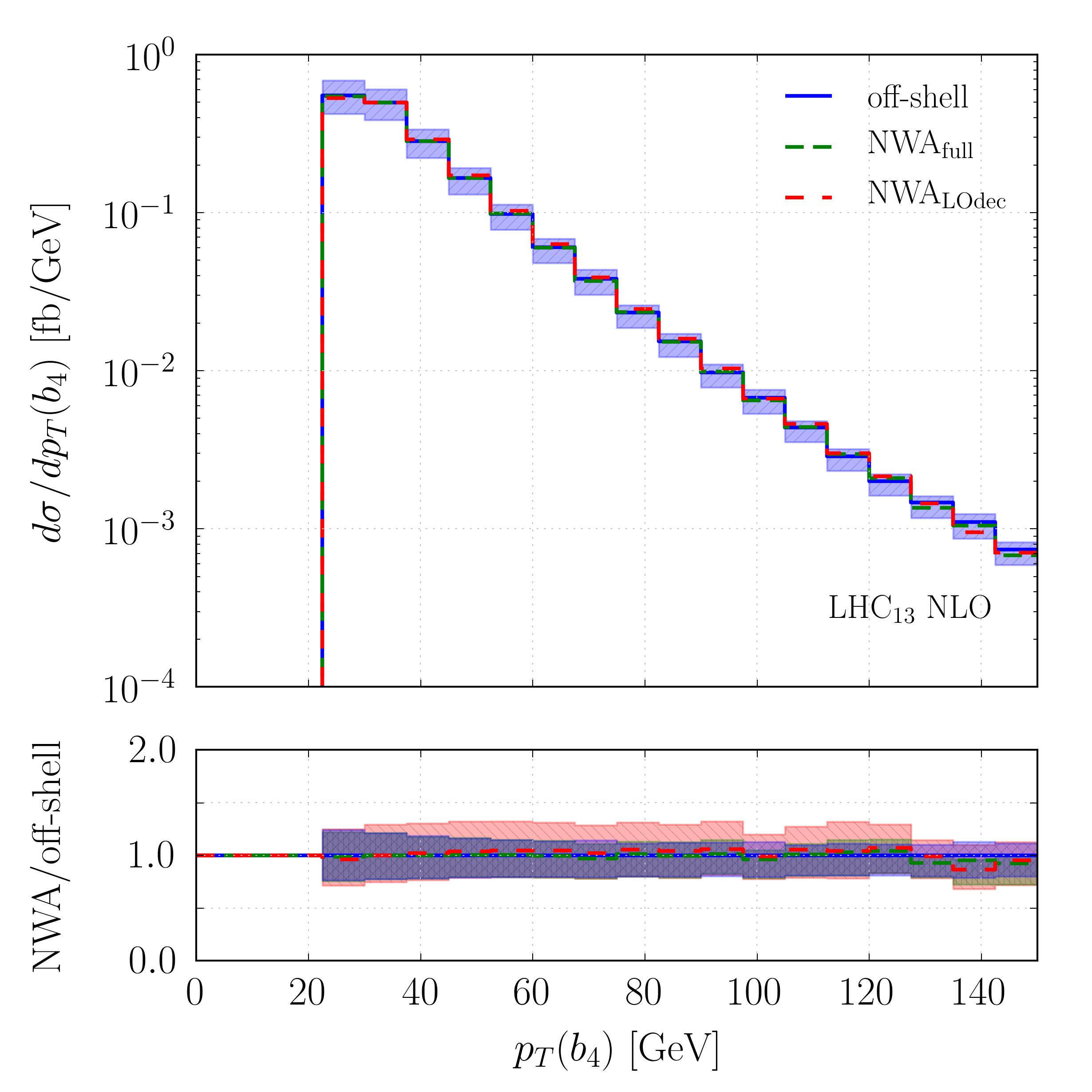} \\
\includegraphics[width=0.49\textwidth]{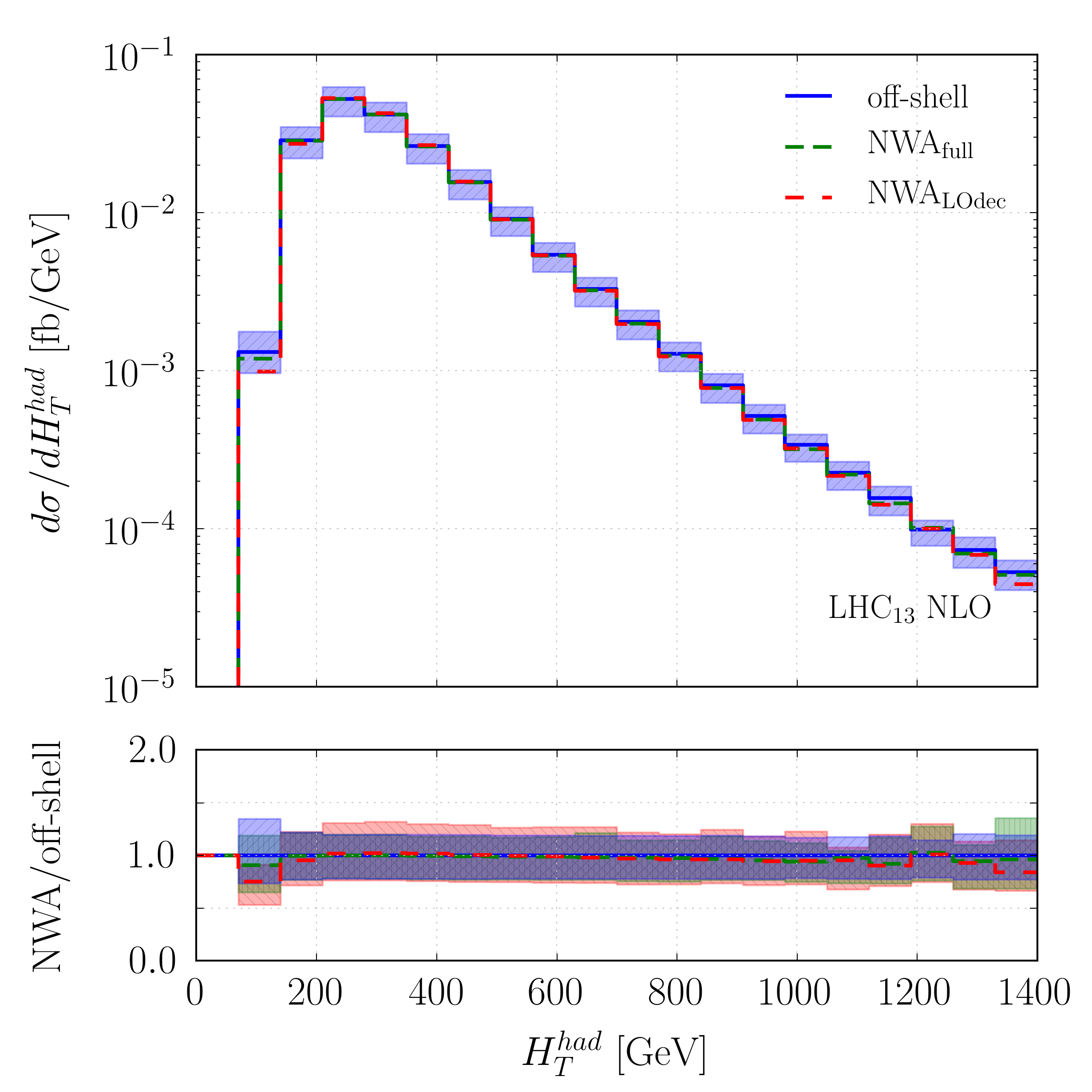}
\includegraphics[width=0.49\textwidth]{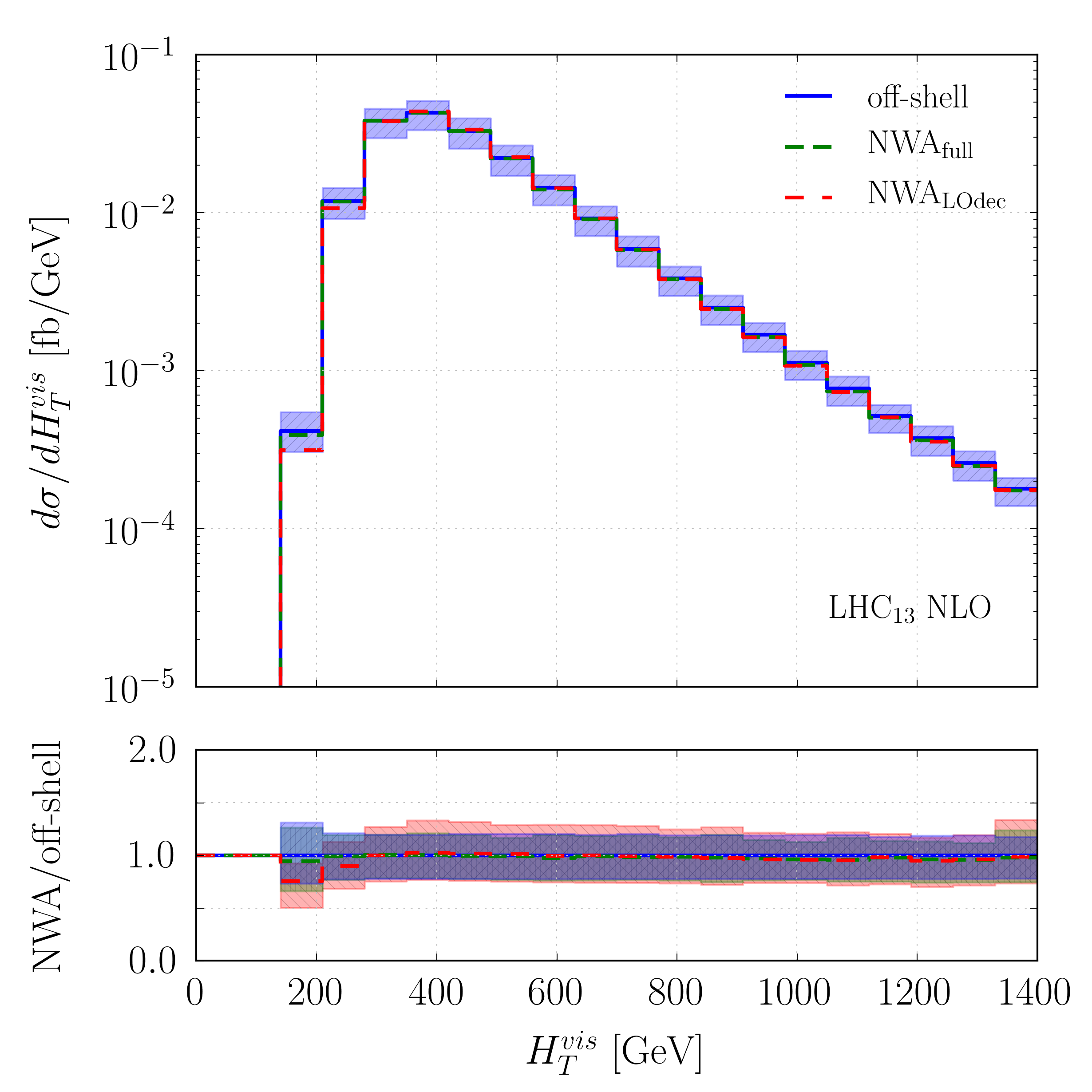}
\caption{\it Differential cross section distributions as a function of
$p_T(b_1)$, $p_T(b_4)$, $H_T^{had}$ and $H_T^{vis}$ at NLO in QCD for
$pp\rightarrow e^+ \nu_e \mu^- \bar{\nu}_\mu b\bar{b} b \bar{b} + X$
at the LHC with $\sqrt{s}=13$ TeV.  The NLO NNPDF3.1 PDF set and the
dynamical scale, $\mu_R = \mu_F = \mu_0=H_T/3$, are employed. In the
upper panel, off-shell results are presented together with their scale
uncertainties while only the central scale predictions are given for
the $\rm NWA_{full}$ and $\rm NWA_{LOdec}$ case. In the lower panel,
the ratios to the full off-shell calculation are shown along with the
scale uncertainties.}
\label{diffxs}
\end{figure}
%

We will now present our theoretical predictions for $pp\rightarrow e^+
\nu_e \mu^- \bar{\nu}_\mu b \bar{b} b \bar{b} + X$ at the differential
level.  We first analyse the impact of the full off-shell effects by means
of systematic comparisons against the NWA. Then we perform a study of
the $b$-jet identification in the full off-shell calculation using the
$\rm NWA_{full}$ results as a benchmark to establish the validity of
our approach.


\subsection{Full off-shell effects}


We have shown in Section~\ref{sec:intxsec} that, at the integrated
fiducial level,  full off-shell effects are rather small in
comparison to the theoretical uncertainties originating from scale
variations.  This turns out to be the same for the majority of
differential cross-section distributions that we have examined.  By
investigating a large set of observables, we found that, for most of
them, we could neither detect noticeable full off-shell effects nor
important contributions stemming from the corrections to top-quark
decays.  As an example, in Figure~\ref{diffxs} we show the transverse
momentum distributions of the hardest and softest $b$ jets, $p_T(b_1)$
and $p_T(b_4)$, as well as two additional observables, $H_T^{vis}$ and
$H_T^{had}$, defined as follows:
\begin{equation}\label{ht-obs}
  H_T^{had} = \sum_{i=1}^4 p_T(b_i),\hspazio\hspazio
  \hspazio\hspazio H_T^{vis} = H_T^{had} + \sum_{i=1}^2 p_T(\ell_i).
\end{equation}
Three different predictions are reported in each plot: the full
off-shell, $\rm NWA_{full}$ and $\rm NWA_{LOdec}$.  In the upper panel
only the scale uncertainty of the full off-shell prediction is
reported.  The lower panel shows the ratio of the various predictions
to the full off-shell one.  Also given are the scale uncertainties.
We can see that the three predictions are in very good agreement for
the central scale values.  Furthermore,
the full off-shell and the $\rm NWA_{full}$ predictions have
comparable theoretical uncertainties.  As we observed for the case of
the integrated fiducial cross section, also here the $\rm NWA_{LOdec}$
exhibits larger scale dependence.

\begin{figure}[t!]
\centering
\includegraphics[width=0.49\textwidth]{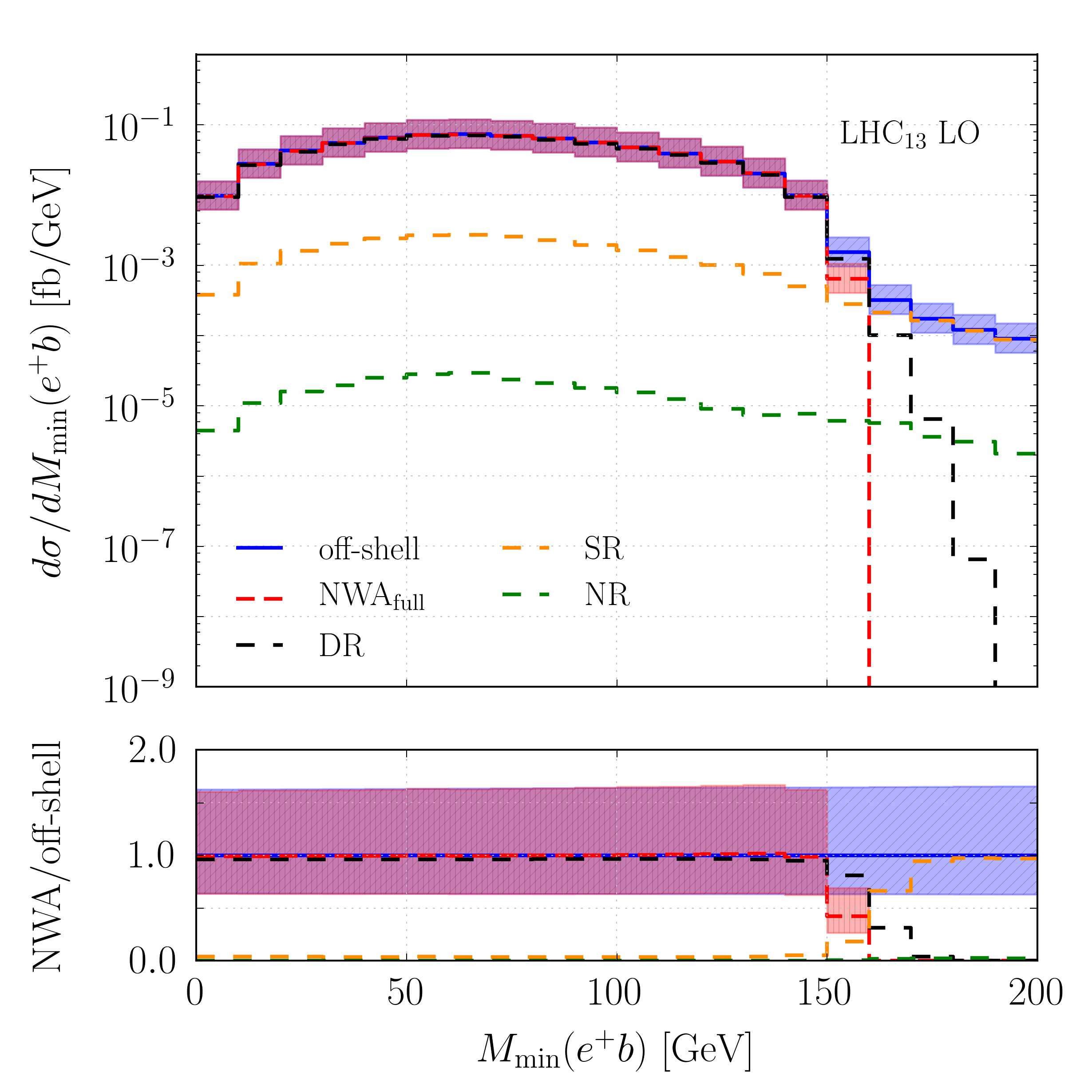}
\includegraphics[width=0.49\textwidth]{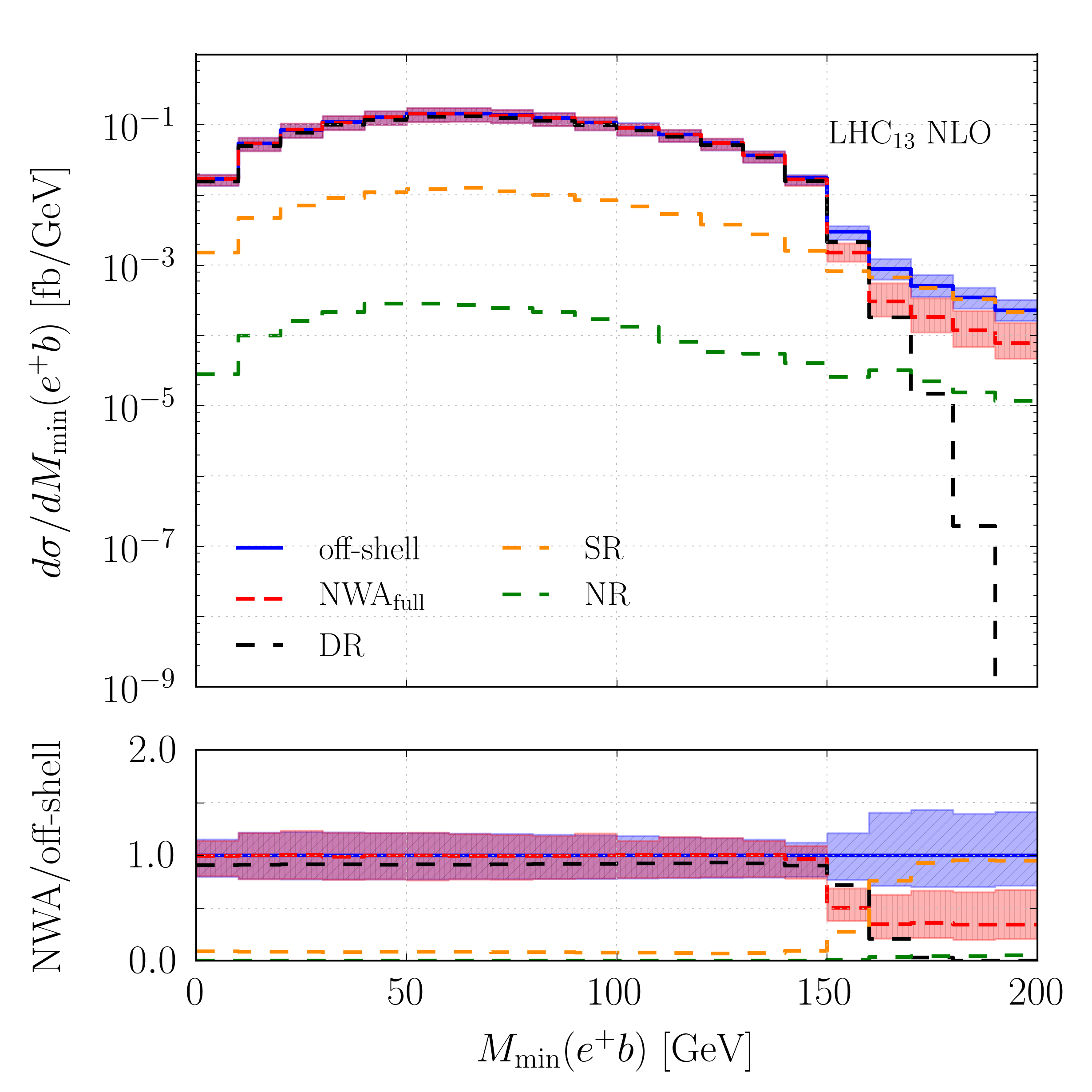}
\caption{\it Differential cross section distributions as a function of
$M_{\rm min}(e^+b)$ at LO (left) and NLO (right) in QCD for
$pp\rightarrow e^+ \nu_e \mu^- \bar{\nu}_\mu b\bar{b} b \bar{b} + X$
at the LHC with $\sqrt{s}=13$ TeV.  The LO and NLO NNPDF3.1 PDF sets
and the dynamical scale, $\mu_R = \mu_F = \mu_0=H_T/3$, are
employed. In the upper panel, full off-shell and $\rm NWA_{full}$
results are presented together with their scale uncertainties.  The
DR, SR and NR contributions to the off-shell prediction are also
displayed in the upper panel.  In the lower panel, the ratios to the
full off-shell calculation are given along with the scale
uncertainties.}
\label{LO-offsh-1}
\end{figure}

We now turn our attention to a different set of observables, that are
related to the top-quark and $W$-boson masses.  We first introduce the
minimal invariant mass of the $b$ jet and the positron
$M_{\rm{min}}(e^+ b)$, defined by
\begin{equation}
M_{\rm{min}}(e^+ b) = \mathrm{min}\left\lbrace M(e^+ b_1),M(e^+
  b_2),M(e^+ b_3),M(e^+ b_4)
\right\rbrace.
\end{equation}
This observable is frequently used for top-quark mass measurements in
$t\bar{t}$ production both by the ATLAS and CMS
collaborations~\cite{ATLAS:2016muw,CMS:2017znf}.  It is furthermore
employed in various BSM analyses in the top-quark
sector~\cite{Haisch:2018djm,Haisch:2018bby,CMS:2018ncu}. Considering
the decay $t \to W^+ b$, and assuming that both $t$ and $W$ are
on-shell, the following kinematical limit applies: $M(e^+b) <
\sqrt{m_t^2-m_W^2}$. Thus, it is not surprising that the NWA at LO
predicts a kinematical edge at $\sqrt{m_t^2-m_W^2} \approx 153$ GeV in
the $M_{\rm min}(e^+b)$ distribution. Off-shell effects at LO as well as
extra-radiation effects starting from NLO are responsible for a
smearing of the kinematical edge, as shown in Figure~\ref{LO-offsh-1}.
%

\begin{figure}[t!]
\centering
\includegraphics[width=0.49\textwidth]{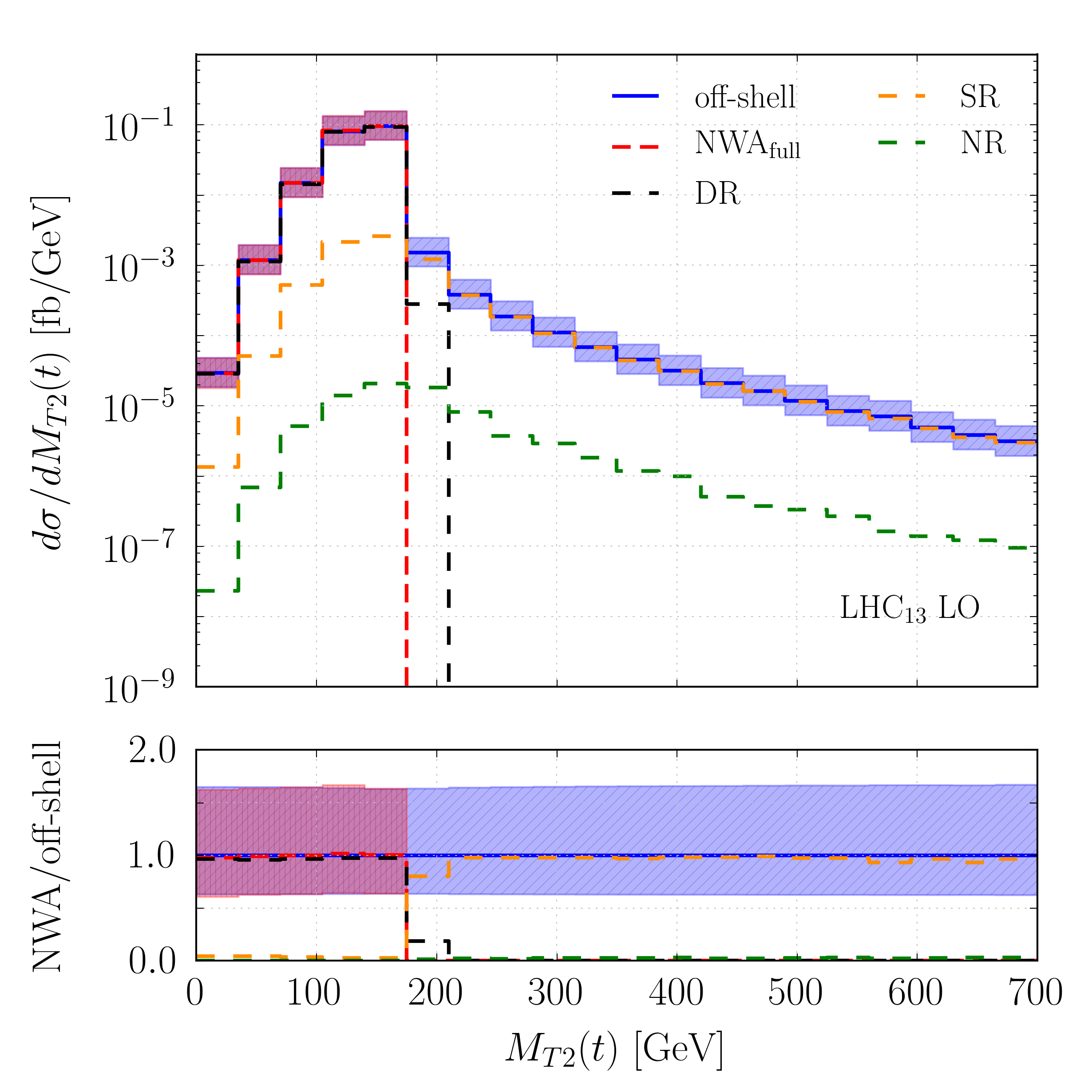}
\includegraphics[width=0.49\textwidth]{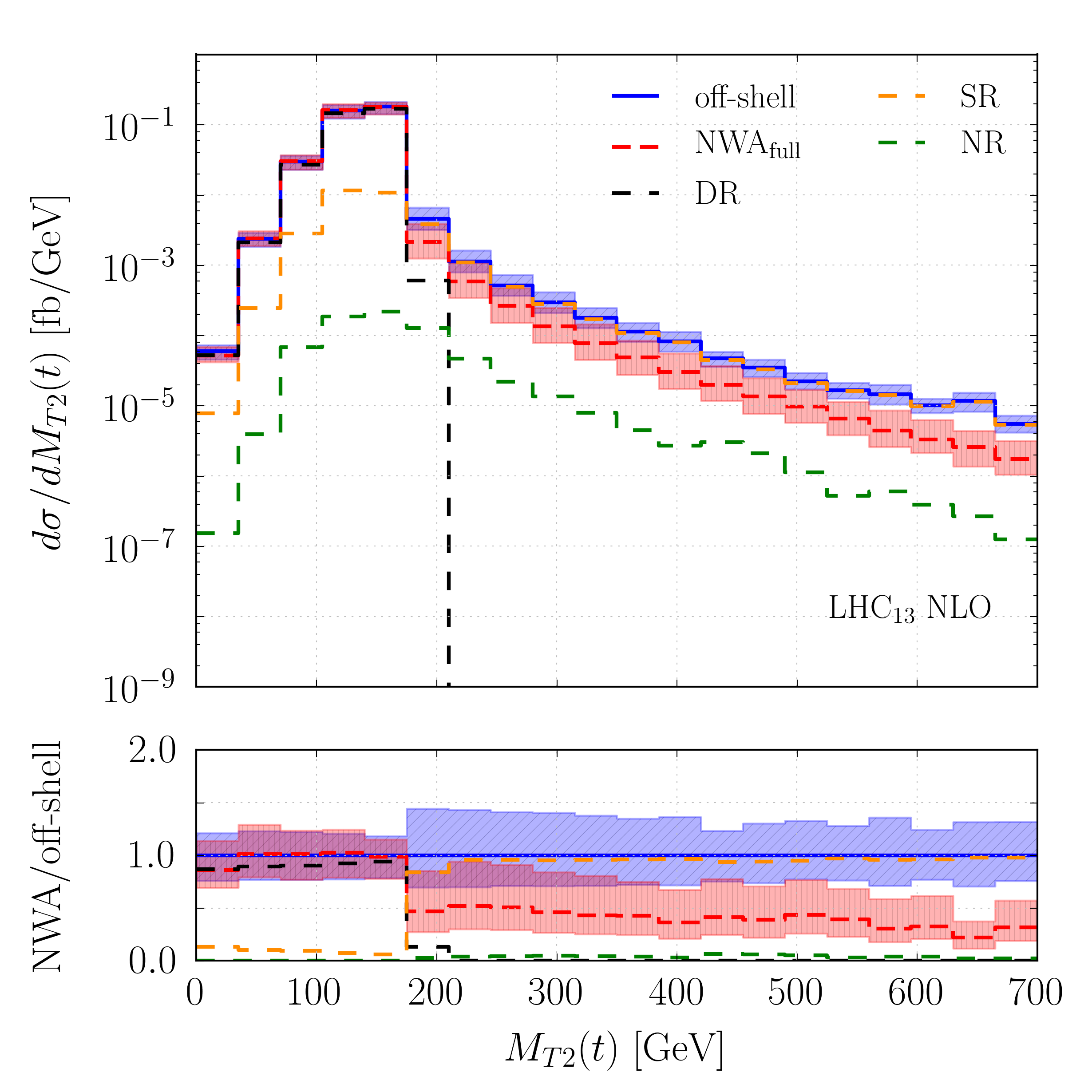}
\includegraphics[width=0.49\textwidth]{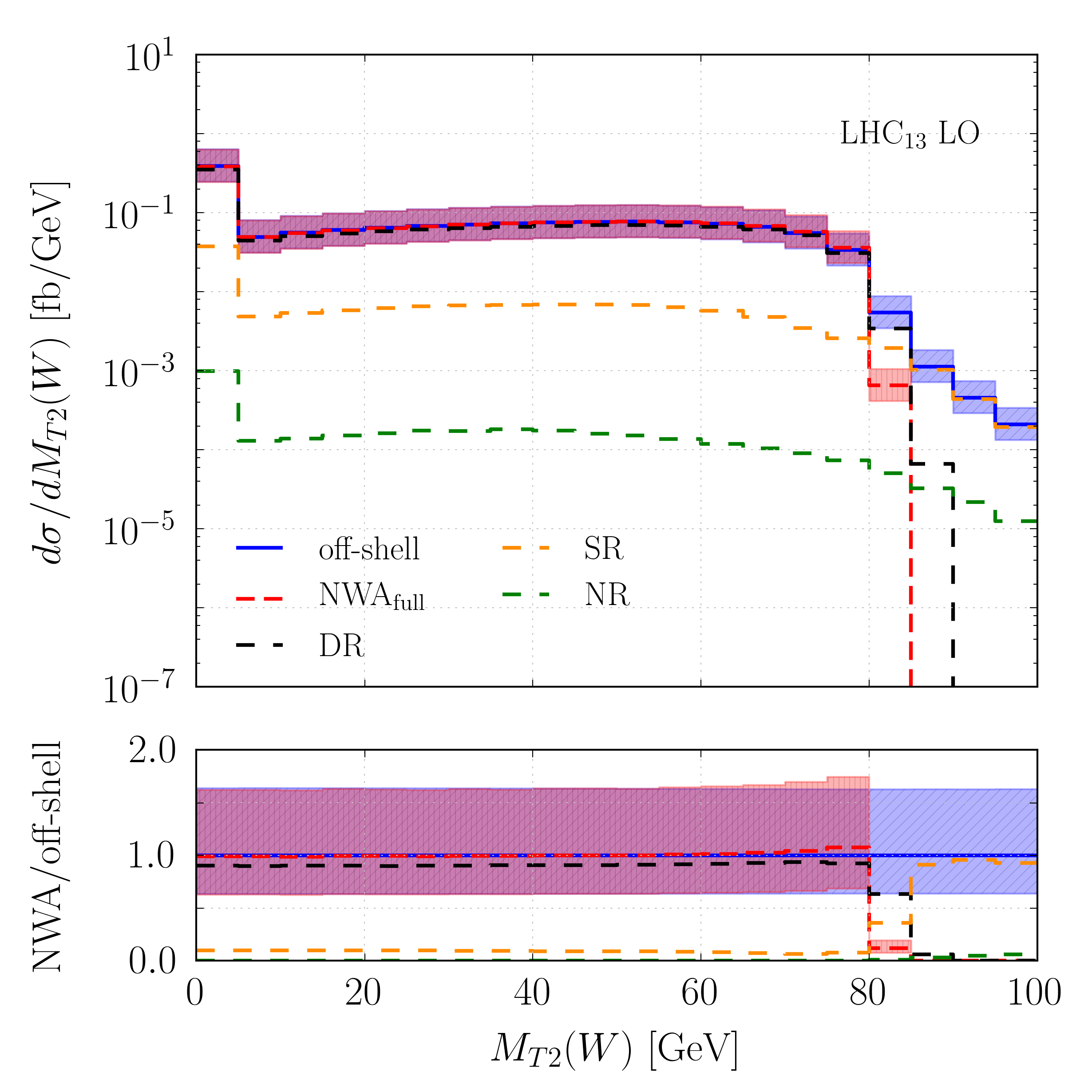}
\includegraphics[width=0.49\textwidth]{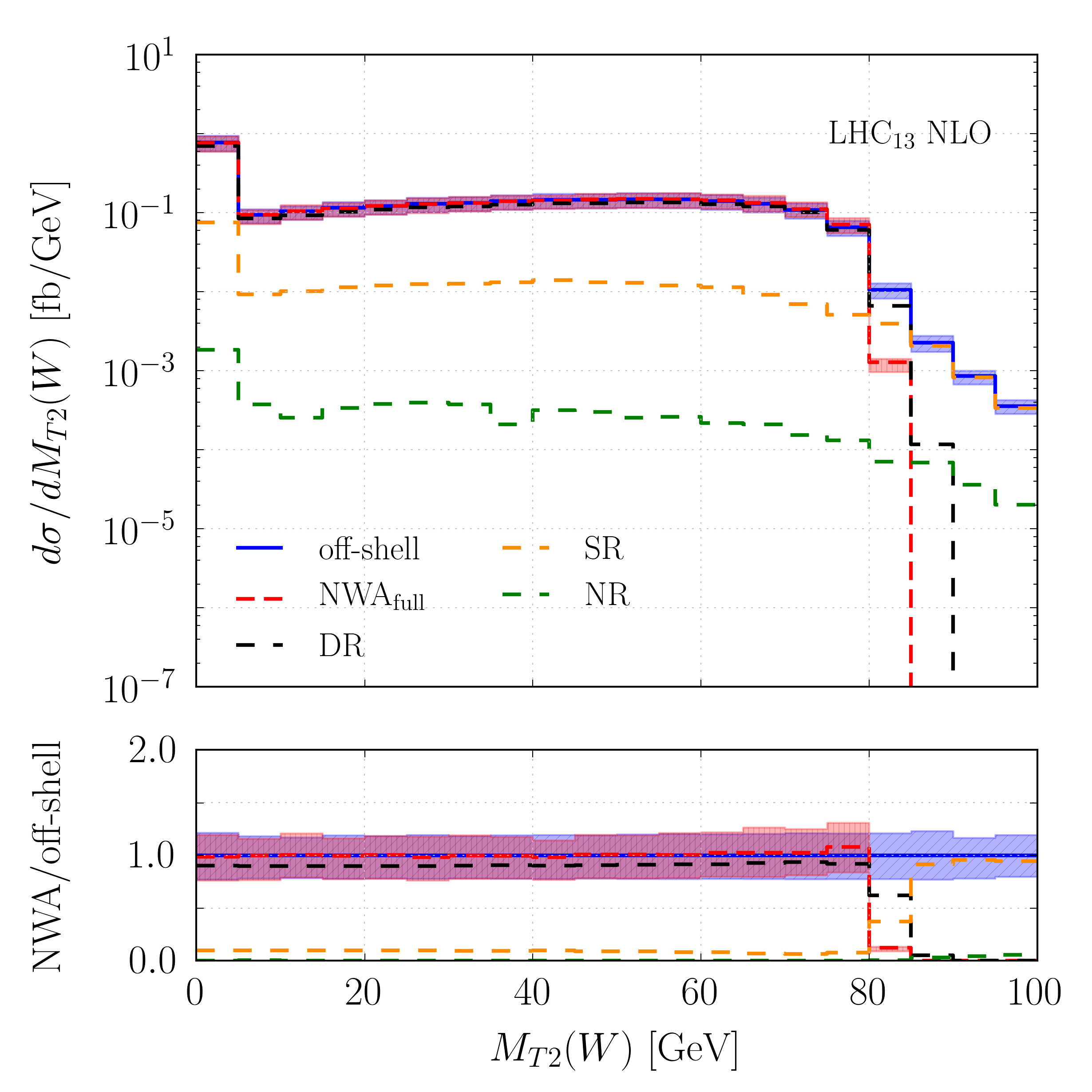}
\caption{\it Differential cross section distributions as a
function of $M_{T2}(t)$ and $M_{T2}(W)$ at LO (left) and NLO (right)
in QCD for $pp\rightarrow e^+ \nu_e \mu^- \bar{\nu}_\mu b\bar{b} b
\bar{b} + X$ at the LHC with $\sqrt{s}=13$ TeV.  The LO and NLO
NNPDF3.1 PDF sets and the dynamical scale, $\mu_R = \mu_F = \mu_0=H_T/3$,
are employed. In the upper panel, full off-shell and $\rm NWA_{full}$
results are presented together with their scale uncertainties. The DR,
SR and NR contributions to the off-shell prediction are also displayed
in the upper panel.  In the lower panel, the ratios to the full
off-shell calculation are given along with the scale
uncertainties.}
\label{LO-offsh-2}
\end{figure}

In the next step, we study two examples of the so-called
\textit{stransverse mass}, indicated as $M_{T2}$, first proposed in
Refs.~\cite{Lester:1999tx,Barr:2003rg}.  To compute these observables,
we adopt the algorithm of Ref.~\cite{Lester:2014yga}.  First, we study
the stransverse mass of the top quark, $M_{T2}(t)$, which is defined as
follows
\begin{equation}
  \label{MT2t}
  M_{T2}(t) = \underset{\sum p_T^{\nu_i}
    =p_T^{miss}}{\text{min}} \bigr[\text{max}\bigr\{M_T^2
  \bigr(p_T(e^+ X_t),p_T(\nu_1)\bigr),M_T^2\bigr(p_T(\mu^-
  X_{\bar{t}}),p_T(\nu_2)
  \bigr)\bigr\}\bigr]\,.
\end{equation}
The quantities $X_t, X_{\bar{t}}$ appearing in Eq.~\eqref{MT2t} indicate
collectively the final-state jets associated to the decays of the
$t,\bar{t}$ quarks according to the selected resonant history. Thus,
at NLO, one has for example $X_t \in (b, bj, bb\bar{b}, bb\bar{b}j )$
and  $X_{\bar{t}} \in (\bar{b}, \bar{b}j,
\bar{b}b\bar{b},\bar{b}b\bar{b}j)$. Let us note that the momenta of
the two final-state neutrinos, $p(\nu_1)$ and $p(\nu_2)$, are unknown
individually as they escape detection, but  they satisfy the following
constraint $p_T(\nu_1+\nu_2) = p_{T}^{miss}$. According to
Eq.~\eqref{MT2t} they are defined to be the pair of momenta which
minimises $M_{T2}(t)$. The quantity $M_{T}^2$ appearing
in Eq.~\eqref{MT2t} is defined as follows
\begin{equation}
    M_T^2\bigr(p_T(\ell X),p_T(\nu)\bigr) = M^2(\ell X) 
    + 2\bigl[ E_T(\ell X) E_T(\nu) - \mathbf{p}_T(\ell X)
    \cdot \mathbf{p}_T(\nu) \bigr],
\end{equation}
where $M(\ell X)$ is the invariant mass of the lepton+jet(s) system,
$E_T(Y)=\sqrt{ \mathbf{p}^2_T(Y)+M(Y)^2}$ and $\mathbf{p}_T = (p_x,
p_y)$ is a two-component vector.  Therefore, $M_T\bigr(p_T(\ell
X),p_T(\nu)\bigr)$ is the transverse
mass of the lepton+jet(s) system for a given choice of the neutrino's
momentum. It is not difficult to see that the $M_{T2} (t)$ observable,
being expressed as a function of transverse masses according to
Eq.~\eqref{MT2t}, exhibits a kinematical edge at $M_{T2} (t) =
m_t$. This observable is typically used in BSM searches to disentangle
the possible signal of  new heavy resonances from the QCD
background~\cite{ATLAS:2014gmw,Haisch:2016gry,Hermann:2021xvs}.  For
the $\rm NWA_{full}$ prediction, we expect at LO an edge at $m_t$ in the
distribution of $M_{T2}(t)$. At NLO, again, the extra radiation will
smear the distribution.  Both LO and NLO predictions for the
$M_{T2}(t)$ observable are depicted in Figure~\ref{LO-offsh-2}.
%
%
%

The second $M_{T2}$ observable  we are going to discuss is the
stransverse mass of the $W$ gauge boson, $M_{T2}(W)$, defined
similarly to the case of the top quark
\begin{equation}
  M_{T2}(W) = \underset{\sum p_T^{\nu_i}
    =p_T^{miss}}{\text{min}} \bigr[\text{max}\bigr
  \{M_T^2\bigr(p_T(e^+),p_T(\nu_1)\bigr),M_T^2\bigr(p_T(\mu^-),p_T(\nu_2)
  \bigr)\bigr\}\bigr],
\end{equation}
where
\begin{equation}
  M_T^2\bigr(p_T(\ell),p_T(\nu)\bigr) =
  M^2(\ell) + 2\bigl( E_T(\ell) E_T(\nu) -
  \mathbf{p}_T(\ell) \cdot \mathbf{p}_T(\nu) \bigr).
\end{equation}
Also in this case, the $M_{T2}(W)$ observable has a kinematical edge
but this time around $m_W$. For $M_{T2}(W)$ we expect a kinematical
edge to be present at both LO and NLO accuracies, since leptonic $W$
boson decays are not affected by QCD radiation. In
Figure~\ref{LO-offsh-2} we also compare the predictions of the $\rm
NWA_{full}$ to the full off-shell calculation for the $M_{T2}(W)$
observable. Uncertainty bands related to scale variation are also
reported.

We note that, for all three cases considered, $M_{\rm min}(e^+b)$,
$M_{T2}(t)$ and $M_{T2}(W)$, the $\rm NWA_{full}$ prediction does not
adequately describe the shape of the differential cross section above
the kinematical edge. The relative differences between the full
off-shell case and the $\rm NWA_{full}$ prediction are
large. Specifically, for $M_{\rm{min}}(e^+ b)$ they are up to $66\%$,
for $M_{T2}(t)$ they are of the order of $80\%$ and for $M_{T2}(W)$
they can be as high as $90\%$. These differences are a consequence of
the lack of single- and non-resonant contributions and interference effects as
well as the of the absence of the finite top-quark and $W$ width effects in
the $\rm NWA_{full}$ prediction. In Figures~\ref{LO-offsh-1} and
\ref{LO-offsh-2} we also report separately the DR, SR and NR
contributions to the full off-shell calculation. We can see that full
off-shell effects become sizeable when the SR contribution starts to
dominate. For the $M_{T2}(W)$ observable, full off-shell effects come
predominantly from the finite-width effects of the $W$ boson as well
as interference effects with the single $W$-resonant contribution.
Therefore, we introduced a region decomposition similar to
Eqs.~\eqref{DR}-\eqref{NR} but applied to the $W$ gauge boson instead
of the top quarks.  For this decomposition we choose
$n=5$ to accommodate the fact that the $W$-boson width is larger than
that of the top quark.

Based on our findings, we draw the conclusion that full off-shell
effects in the $t\bar{t}b\bar{b}$ process are rather small for the
majority of observables used in standard analyses. We have shown,
however, that in the case of observables with a kinematical
edge, which are often used in BSM searches, such effects are very
significant. The $\rm NWA_{full}$ prediction simply fails to properly  describe
the phase-space regions above the kinematical edge.


\subsection{$b$-jet identification}
\label{bjet}


%
\begin{figure}[t!]
\centering
\includegraphics[width=0.49\textwidth]{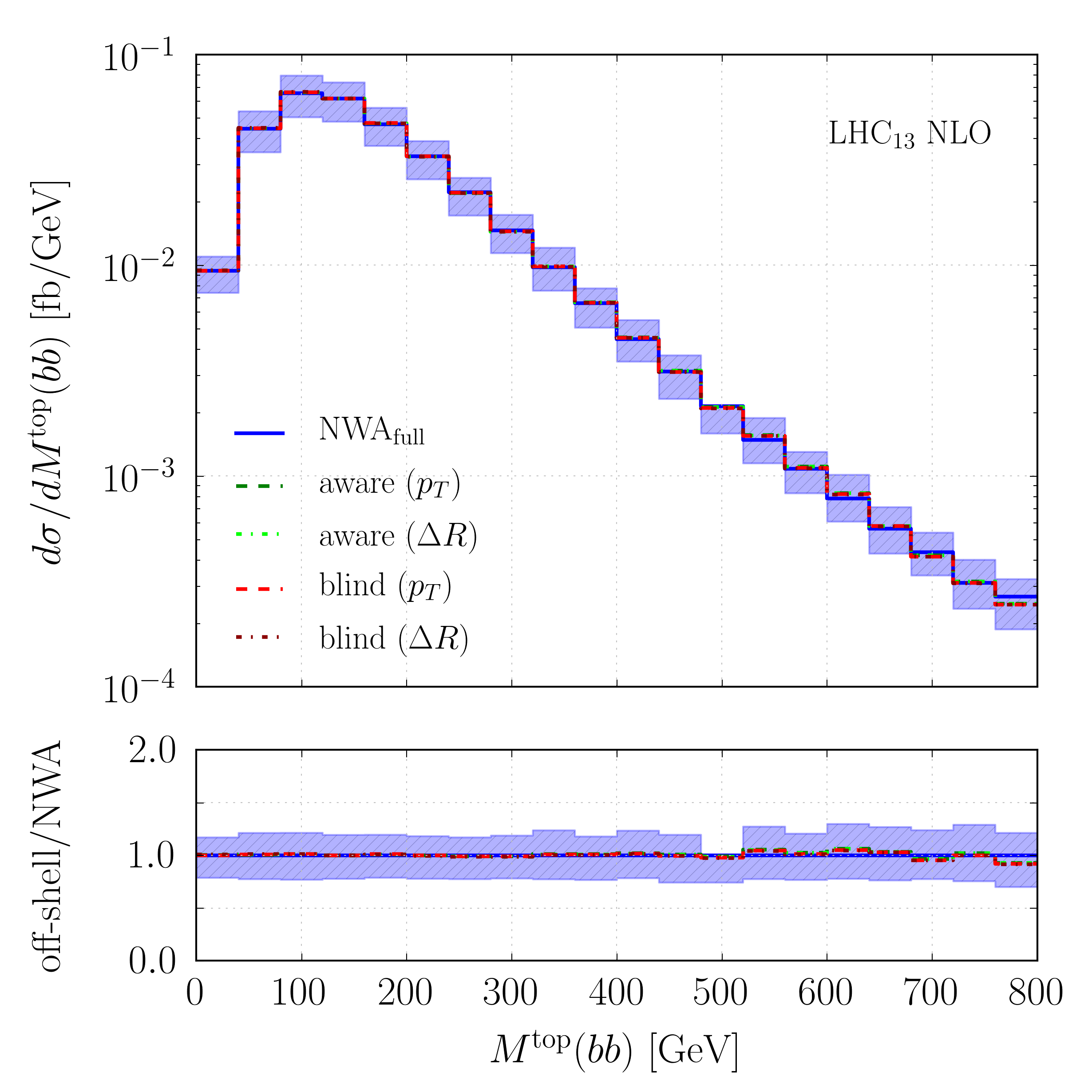}
\includegraphics[width=0.49\textwidth]{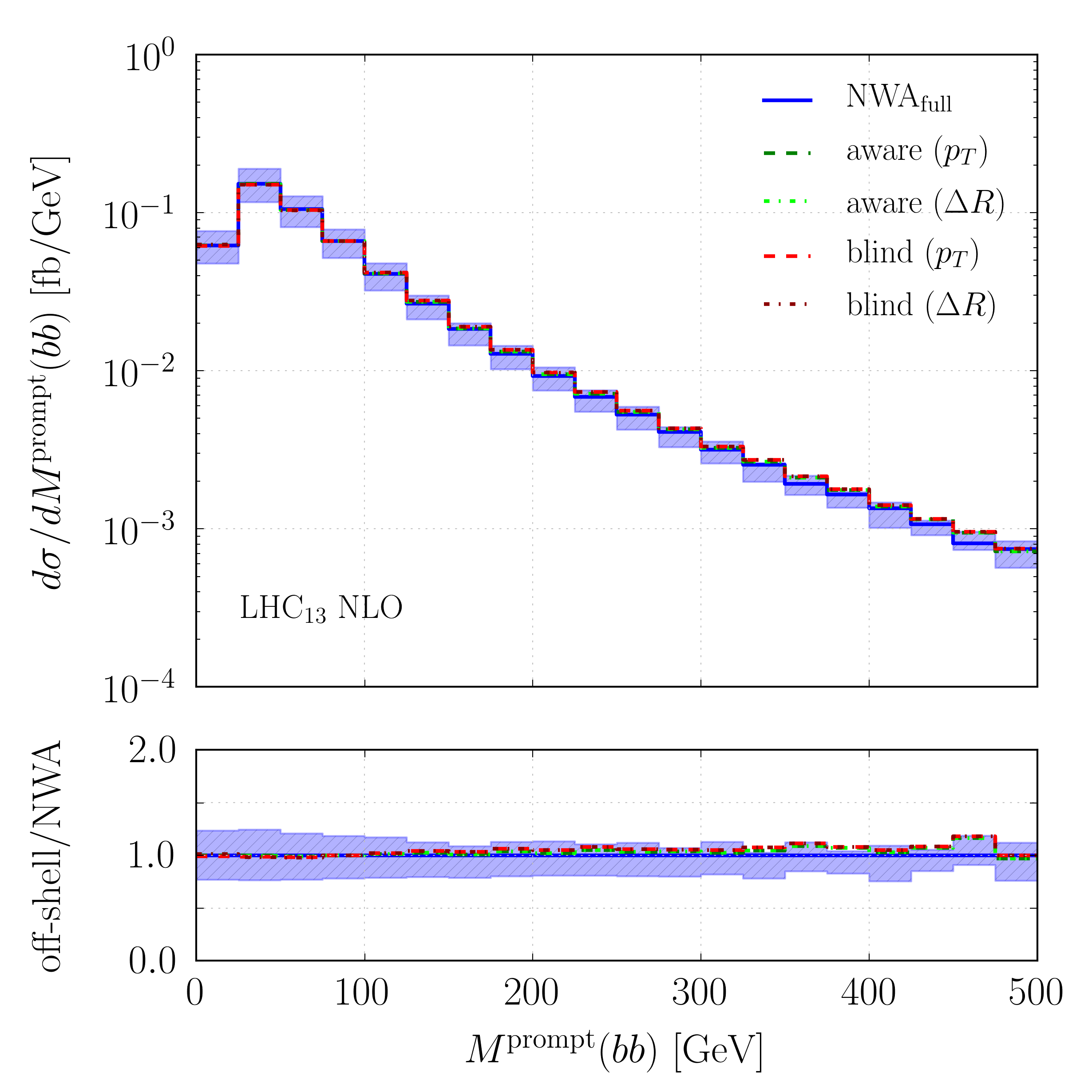}
\caption{\it Differential cross section distributions as a function of
observables describing the $b\bar{b}$ pair coming from top-quark
decays and the prompt $b\bar{b}$ pair at NLO in QCD for $pp\rightarrow
e^+ \nu_e \mu^- \bar{\nu}_\mu b\bar{b} b \bar{b} + X$ at the LHC with
$\sqrt{s}=13$ TeV.  The NLO NNPDF3.1 PDF set and the dynamical scale,
$\mu_R = \mu_F = \mu_0=H_T/3$, are employed. In the upper panel, the
$\rm NWA_{full}$ result is presented together with its scale
uncertainty while only the central scale predictions are shown for the
full off-shell predictions. The latter are obtained using the
charge-aware and charge-blind labellings in combination with different
choices for the discriminator ($p_T$ and $\Delta R$). In the lower
panel, the ratios to the $\rm NWA_{full}$ calculation are given along
with the scale uncertainty of the $\rm NWA_{full}$ case.}
\label{lab-nlo-1}
\end{figure}
\begin{figure}[t!]
\centering
\includegraphics[width=0.49\textwidth]{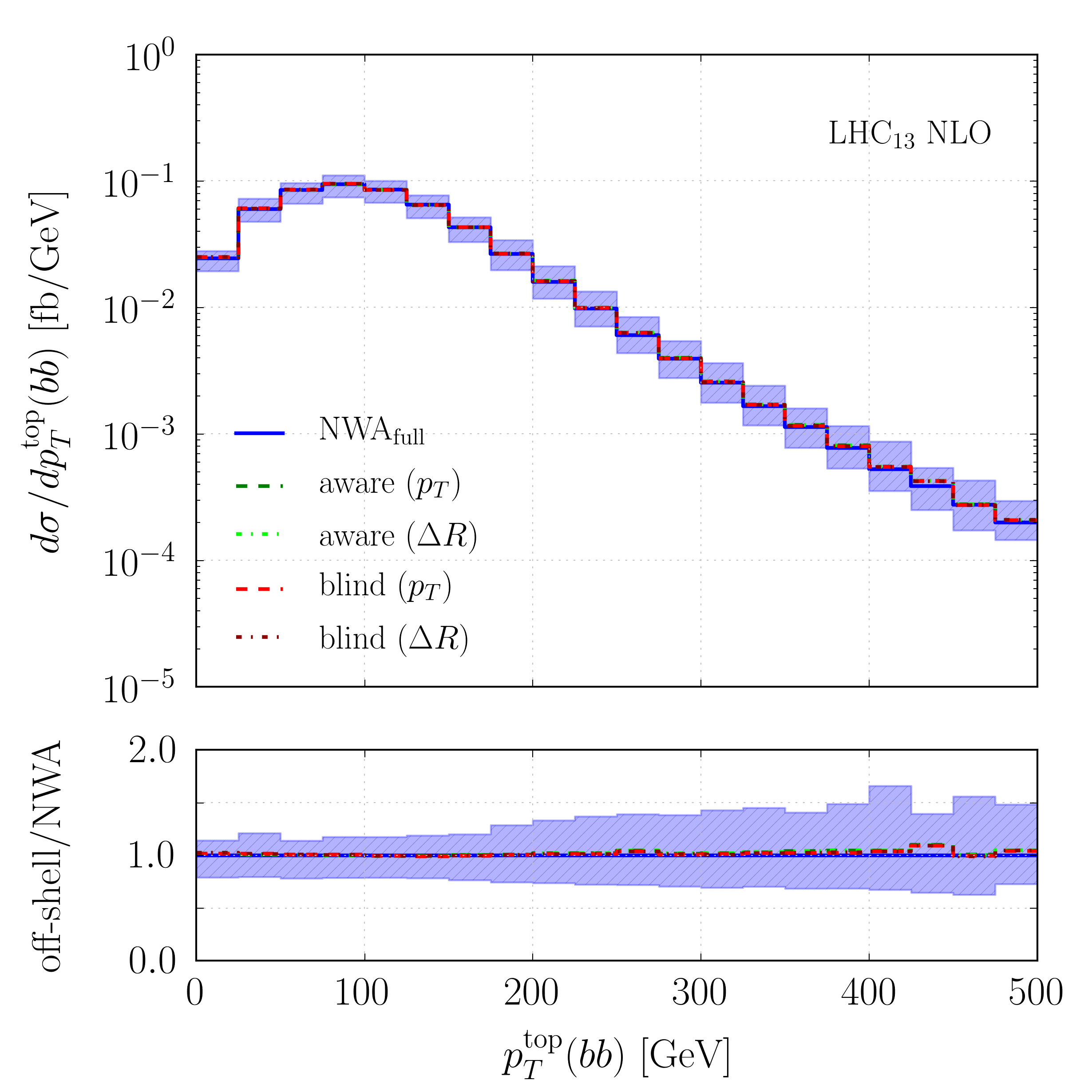}
\includegraphics[width=0.49\textwidth]{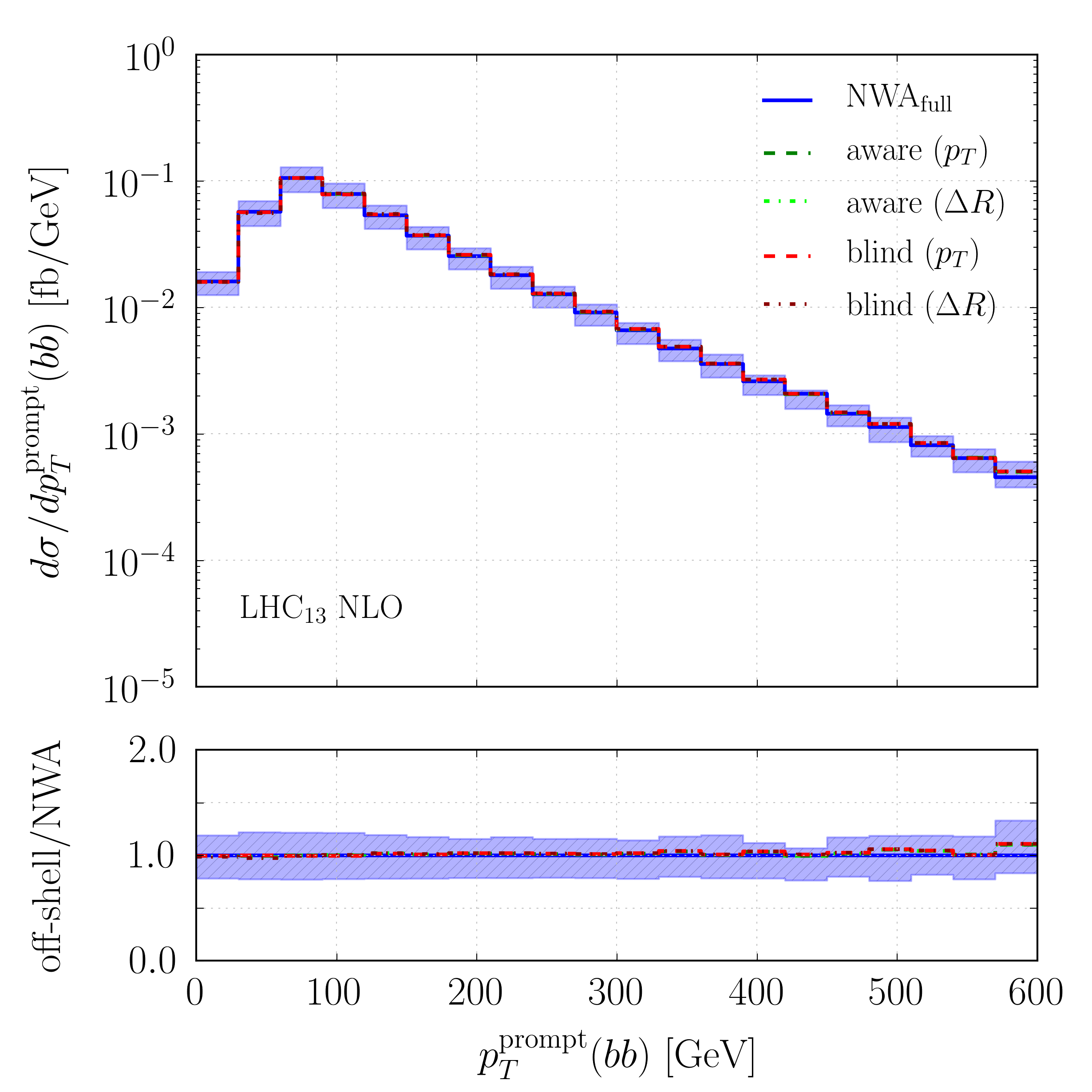}
\includegraphics[width=0.49\textwidth]{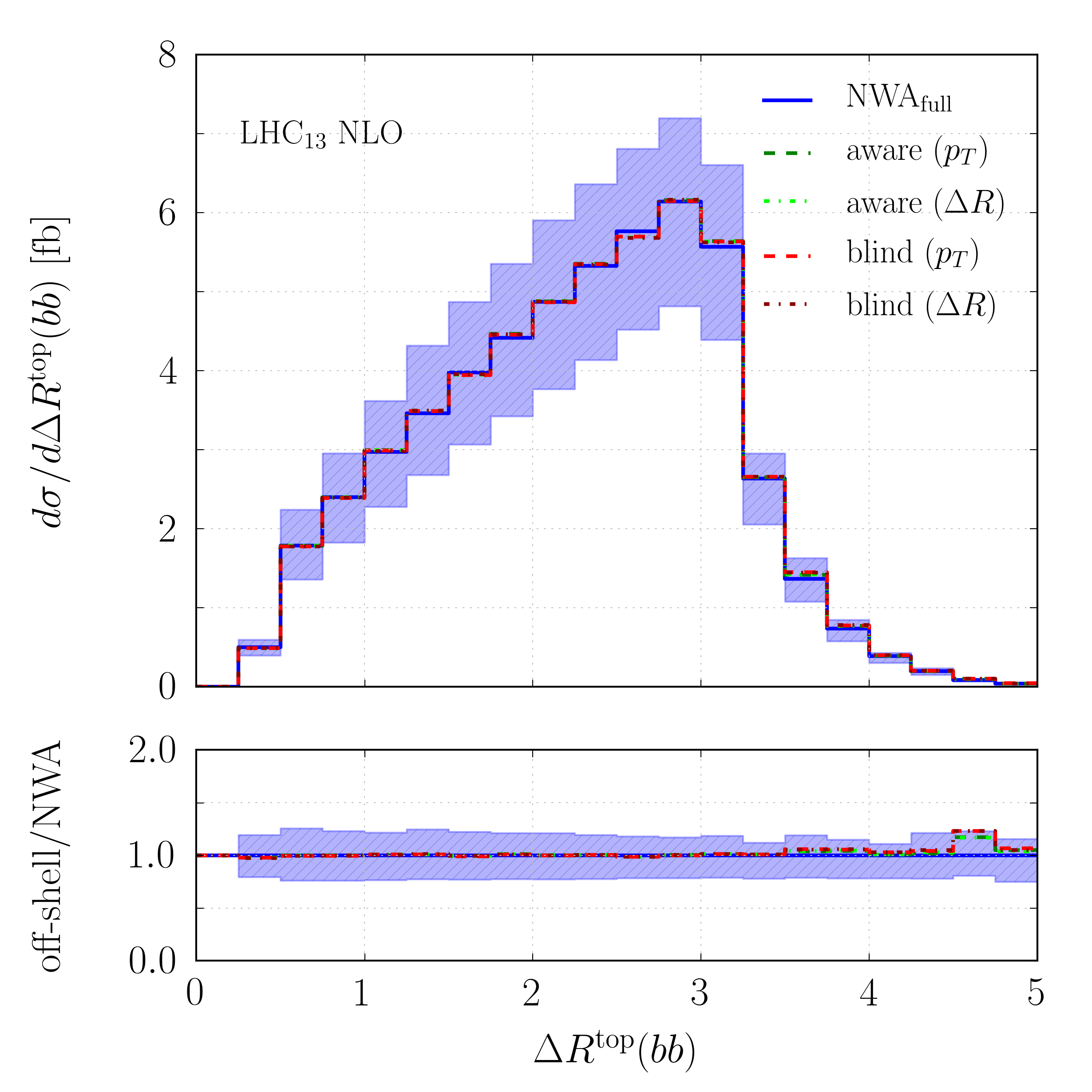}
\includegraphics[width=0.49\textwidth]{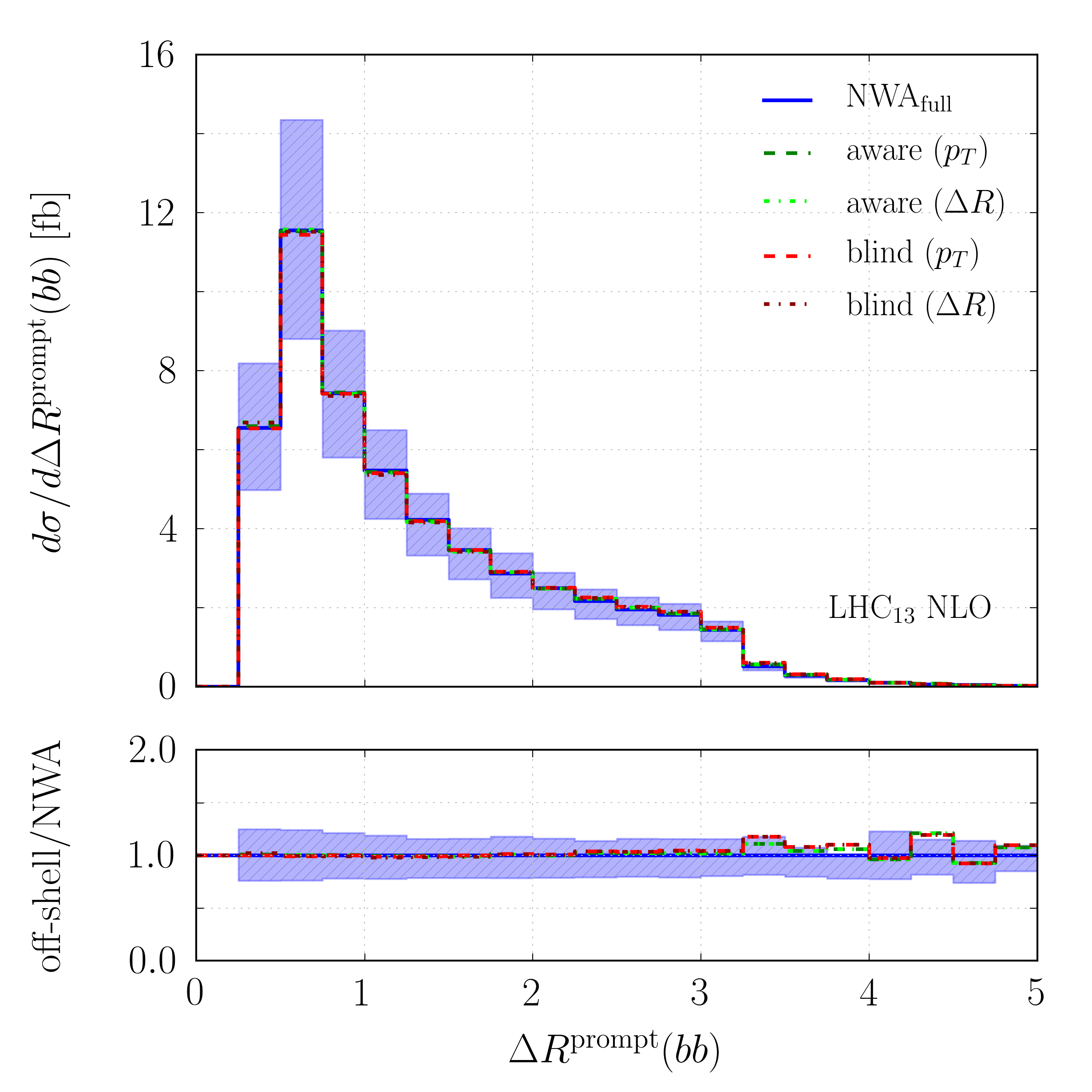}
\caption{\it Differential cross section distributions as a function of
observables describing the $b\bar{b}$ pair coming from top-quark
decays and the prompt $b\bar{b}$ pair at NLO in QCD for $pp\rightarrow
e^+ \nu_e \mu^- \bar{\nu}_\mu b\bar{b} b \bar{b} + X$ at the LHC with
$\sqrt{s}=13$ TeV.  The NLO NNPDF3.1 PDF set and the dynamical scale,
$\mu_R = \mu_F = \mu_0=H_T/3$, are employed. In the upper panel, the $\rm
NWA_{full}$ result is presented together with its scale uncertainty
while only the central scale predictions are shown for the full
off-shell predictions. The latter are obtained using the charge-aware and
charge-blind labellings in combination with different choices for the
discriminator ($p_T$ and $\Delta R$). In the lower panel, the ratios
to the $\rm NWA_{full}$ calculation are given along with the scale
uncertainty of the $\rm NWA_{full}$ case.}
\label{lab-nlo-2}
\end{figure}
\begin{figure}[t]
\centering
\includegraphics[width=0.45\textwidth]{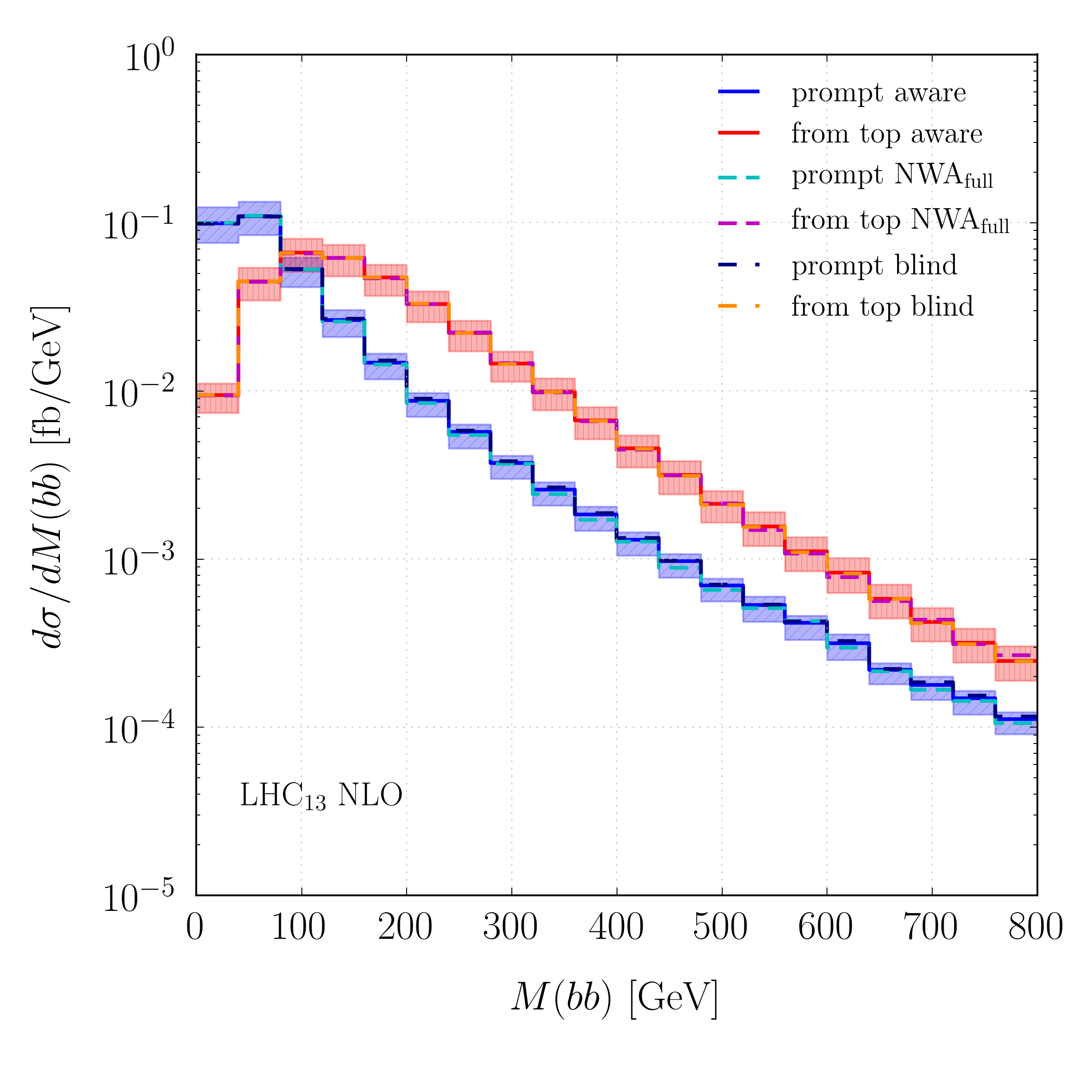}
\includegraphics[width=0.45\textwidth]{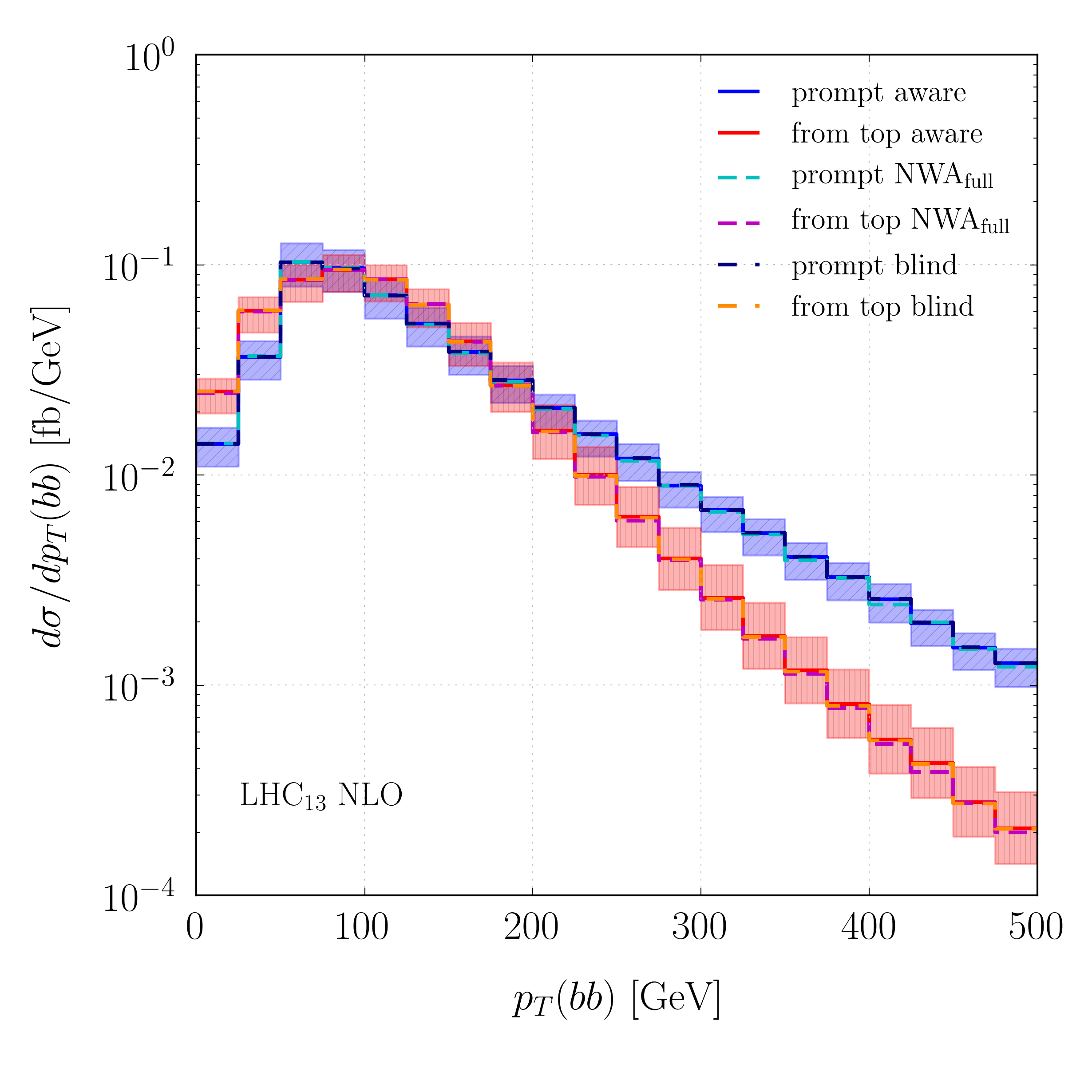}
\includegraphics[width=0.45\textwidth]{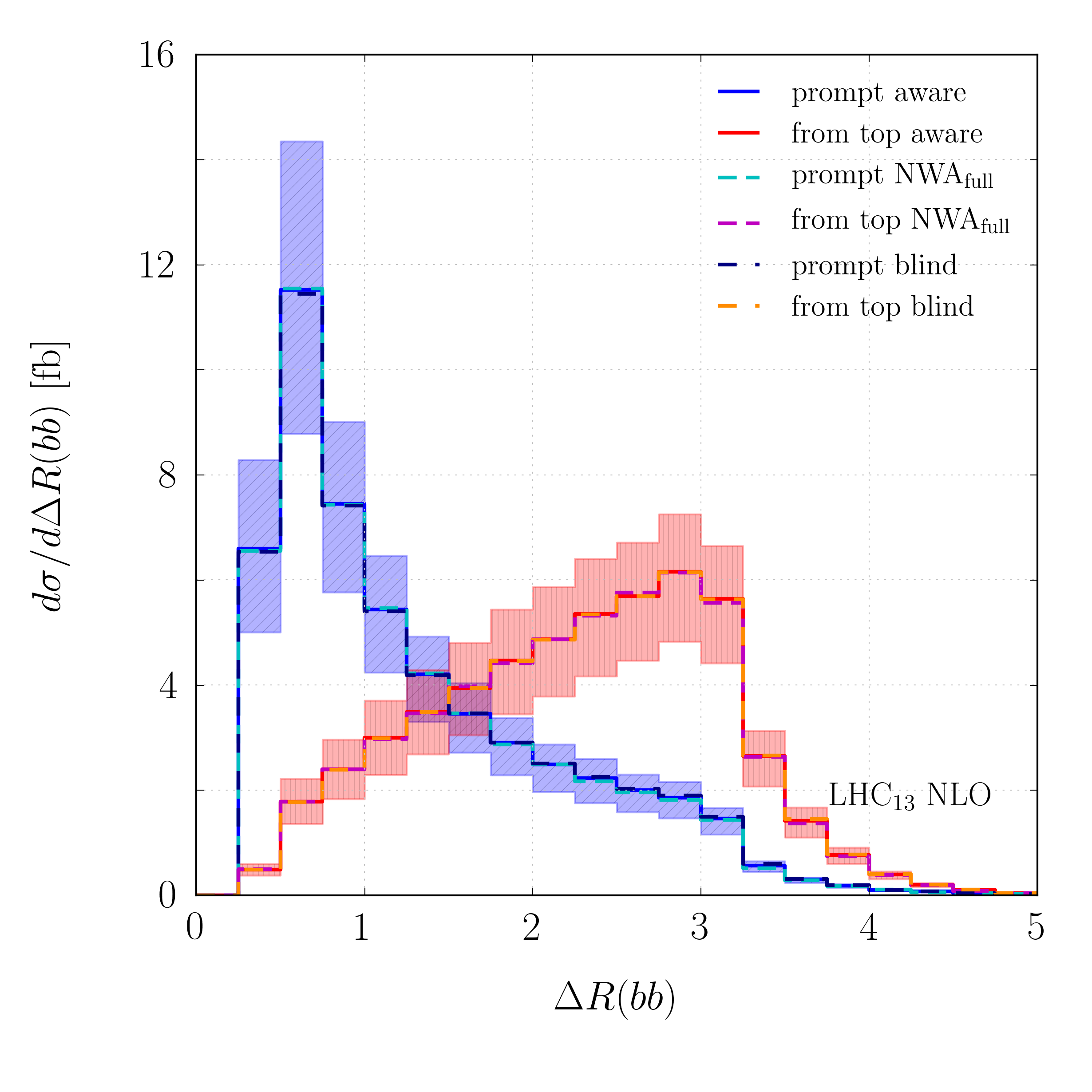}
\caption{\it Differential cross section distributions as a function of
observables describing the $b\bar{b}$ pair coming from top-quark
decays and the prompt $b\bar{b}$ pair at NLO in QCD for $pp\rightarrow
e^+ \nu_e \mu^- \bar{\nu}_\mu b\bar{b} b \bar{b} + X$ at the LHC with
$\sqrt{s}=13$ TeV.  The NLO NNPDF3.1 PDF set and the dynamical scale,
$\mu_R = \mu_F = \mu_0=H_T/3$, are employed. The distributions for the two
$b\bar{b}$ pairs are reported in the same plot. The $\rm NWA_{full}$ results
are presented together with the full off-shell predictions, which are
obtained using the charge-aware and charge-blind labellings with the $p_T$
discriminator. The scale uncertainty bands for the charge-aware
predictions are also displayed.}
\label{comparison}
\end{figure}

Applying the techniques for $b$-jet identification described in
Section~\ref{sec:toprecon}, we can study a number of observables
related to the kinematics of the $b$ jets associated with top-quark
decays as well as of those labelled as prompt $b$ jets. We focus in
particular on the invariant mass, transverse momentum and  $\Delta
R$ separation of the $b\bar{b}$ pairs split according to their
labelling. We denote this set of observables $M^{\text{top}}(bb)$,
$p_T^{\text{top}}(bb)$, $\Delta R^{\text{top}}(bb)$ and
$M^{\text{prompt}}(bb)$, $p_T^{\text{prompt}}(bb)$, $\Delta
R^{\text{prompt}}(bb)$ respectively. Mindful of the results presented
in Section~\ref{sec:toprecon}, we compare the full off-shell
predictions to the $\rm NWA_{full}$ result taken as reference, to see how
good our $b$-jet labelling prescription works.  We report our results
in Figures~\ref{lab-nlo-1} and \ref{lab-nlo-2}. In the upper
panel of the plots the differential cross-section distributions are
reported.  We also display the uncertainty band for the reference
$\rm NWA_{full}$ case.  The lower panel reports our full off-shell
predictions, as obtained with different labelling approaches,
normalized to the $\rm NWA_{full}$ prediction. We can see that the
charge-aware and charge-blind labellings yield very similar
predictions, which suggests that the charge information is not very
crucial in our $b$-jet identification procedure.  Moreover, we cannot
see any differences between the two discriminator prescriptions. This
is mainly due to the fact that the prompt $b$-jet pair is
predominantly produced in the production stage, where the
discriminator does not play any role. The predictions based on all
approaches adopted for the identification of $b$ jets are in very good
agreement with the $\rm NWA_{full}$ reference, suggesting that our
labelling prescription performs very well in distinguishing between
the $b$ jets from top-quark decays and the prompt ones.

To better outline kinematical differences between $b\bar{b}$ pairs belonging
to the prompt and non-prompt categories, in Figure~\ref{comparison}
we display afresh and in the same plot NLO differential cross-section
distributions as a function of $M(bb)$, $p_T(bb)$ and $\Delta
R(bb)$. We present various predictions obtained for the $\rm
NWA_{full}$ and full off-shell case.  For the latter, we only report
the results using the $p_T$ discriminator.  We also show the
uncertainty bands as obtained from the scale dependence of the
full off-shell prediction in the charge-aware case. From the $M(bb)$
distribution we can see that the spectrum of $M^{\rm prompt}(bb)$ is
softer than $M^{\rm top}(bb)$. For the $p_T(bb)$ distribution we find
that $p_T^{\rm top}(bb)$ decreases much faster than $p_T^{\rm
prompt}(bb)$. Finally, for the $\Delta R(bb)$ distribution we can
observe that the prompt $b$ jets are more likely to have a smaller angular
separation, while the $b$ jets from top-quark decays are usually
produced in back-to-back configurations.  These results are consistent
with expectations dictated by the dynamics of the $g\to b\bar{b}$
splitting and top-quark decays. We want to point out that one could
naively think that the $b\bar{b}$ pair made of the two highest-$p_T$
$b$ jets is expected to originate from the $t\bar{t}$ pair. However,
by comparing Figure~\ref{comparison} with the corresponding plots in
Ref.~\cite{Bevilacqua:2021cit}, we find that this naive prescription
is not very accurate.

It is worth to mention that several studies of the $t\bar{t}b\bar{b}$
process, that provide predictions for $M^{\rm prompt}(bb)$, $p_T^{\rm
prompt}(bb)$ and $\Delta R^{\rm prompt}(bb)$ distributions at NLO QCD
accuracy, are available in the literature. In the first studies of
this kind, see
e.g. Refs.~\cite{Bevilacqua:2009zn,Bredenstein:2010rs,Bevilacqua:2014qfa,
LHCHiggsCrossSectionWorkingGroup:2016ypw}, top quarks were treated as
stable particles, which makes the identification of prompt $b$ jets
straightforward. We can use these results as important benchmarks to
assess the performance of our $b$-jet labelling prescription. The
issue of $b$-jet identification in the $t\bar{t}b\bar{b}$ process has
also been studied more recently using deep neural network techniques,
see e.g. Refs. \cite{Choi:2019lyt,Jang:2021eph,Cavallini:2021vot}. The
results presented in this Section are fully consistent with the
findings from earlier studies mentioned above. Very good agreement is
found in the shape of the distributions as well as in the location of
the peaks. This confirms even further that our simple $b$-jet
labelling prescription works very well.


\section{Summary}
\label{sec:summ}


In this paper we investigated the impact of the full off-shell effects in
the $pp\to e^+ \nu_e \mu^- \bar{\nu}_\mu b\bar{b}b\bar{b}+X$ process
at NLO in QCD by presenting a comprehensive comparison between the
full off-shell calculation and the NWA predictions.  This is the first
time that the full NWA predictions at NLO in QCD, including
higher-order corrections to both production and decay stages as well
as NLO spin correlations, have been presented for the
$t\bar{t}b\bar{b}$ process in the dilepton channel. Let us stress that
dropping the approximation of on-shell top quarks and $W$ gauge bosons
in the calculation of the $e^+ \nu_e \,\mu^- \bar{\nu}_\mu \,b\bar{b}
\,b\bar{b} +X$ final state not only increases dramatically the
complexity of the calculation but also represents a qualitatively
different level of theoretical description.

At the integrated fiducial level, we found that the full off-shell effects are
very small, of the order of $0.5\%$, which is compatible with the expected size
$\mathcal{O}(\Gamma_t/m_t)$. Moreover, the scale uncertainties for both the
$\rm NWA_{full}$ and the full off-shell calculation are the same and at the
level of $22\%$.  We observed that the NLO QCD corrections to top-quark decays
are negative and amount to $6.6\%$. We also observed that the scale uncertainty
is slightly larger if top-quark decays are treated with LO accuracy. Indeed, it
increases up to $ 29\%$.  Additionally, we noticed that the contribution of
prompt $b$ jets originating from top-quark decays is only of the order of $1\%$
in the setup of our analysis.  At the differential level, for a large number of
observables, the full off-shell effects are small for the energy ranges we
considered.  Not only the $\rm NWA_{full}$ prediction agrees very well with the
full off-shell result, but also the theoretical uncertainties due to missing
higher-order corrections are very similar. The scale uncertainty bands for the
$\rm NWA_{LOdec}$ case, however, are generally larger. Nevertheless, all
predictions are in good agreement within the corresponding theoretical
uncertainties. Our findings suggest that $\rm NWA_{full}$ is a good
approximation for the $pp\to e^+ \nu_e \mu^- \bar{\nu}_\mu b\bar{b}b\bar{b}+X$
process for a large number of observables. However, substantial full off-shell
effects are present for observables with kinematic edges. These effects
originate from single- and non-resonant contributions, interference effects as
well as from top-quarks and $W$-bosons finite-width corrections.

In this paper we also provided a prescription to distinguish prompt
$b$ jets from $b$ jets that come from top-quark decays. The importance
of this study is related to the fact that the $pp\to t\bar{t}b\bar{b}$
process has the same final state as $pp\to t\bar{t}H$ production when
the Higgs boson decays into $H\to b\bar{b}$. This particular decay
chain is crucial for the direct measurement of the top-quark-Higgs and
bottom-quark-Higgs Yukawa couplings. Complementary to the widely used
deep neural network approaches adopted in the literature, we proposed
a kinematic-based prescription where we looked for the most likely
resonant history which might have generated the corresponding final
state. We applied our prescription to the full off-shell calculation
using the NWA results as a benchmark.  By studying the $M(bb)$,
$p_T(bb)$ and $\Delta R(bb)$ distributions of the $b$ jets we found
that our prescription is able to capture the different origin of the
$b$ jets which appear in the final state.


\section*{Acknowledgments}
The work of H.Y.B, M.L. and M.W.  was supported by the Deutsche
Forschungsgemeinschaft (DFG) under the following grants 396021762 $-$ TRR 257:
{\it P3H - Particle Physics Phenomenology after the Higgs Discovery} and
400140256 - GRK 2497: {\it The physics of the heaviest particles at the Large
Hardon   Collider}. Support by a grant of the Bundesministerium f\"ur Bildung
und Forschung (BMBF) is additionally acknowledged.

The research of G.B. is supported by grant K 125105 of the National Research,
Development and Innovation Office in Hungary.

H.B.H. has received funding from the European Research Council (ERC) under the
European Union's Horizon 2020 Research and Innovation Programme (grant
agreement no. 683211). Furthermore, the work of H.B.H has been partially
supported by STFC consolidated HEP theory grant ST/T000694/1

The work of M.K. was supported in part by the U.S. Department of Energy under
grant DE-SC0010102.

Simulations were performed with computing resources granted by RWTH Aachen
University under projects {\tt rwth0414}, {\tt rwth0878}  and
{\tt rwth0846}.


\providecommand{\href}[2]{#2}\begingroup\raggedright\endgroup


\begin{thebibliography}{10}

\bibitem{ATLAS:2012yve}
{\bf ATLAS} Collaboration, G.~Aad et~al., {\it {Observation of a new particle
  in the search for the Standard Model Higgs boson with the ATLAS detector at
  the LHC}},  Phys. Lett.  {\bf B 716} (2012) 1
  [\href{http://arxiv.org/abs/1207.7214}{{\tt arXiv:1207.7214}}].

\bibitem{CMS:2012qbp}
{\bf CMS} Collaboration, S.~Chatrchyan et~al., {\it {Observation of a New Boson
  at a Mass of 125 GeV with the CMS Experiment at the LHC}},  Phys. Lett.
  {\bf B 716} (2012) 30 [\href{http://arxiv.org/abs/1207.7235}{{\tt
  arXiv:1207.7235}}].

\bibitem{Glashow:1961tr}
S.~L. Glashow, {\it {Partial Symmetries of Weak Interactions}},   Nucl.
  Phys. {\bf 22} (1961) 579.

\bibitem{Weinberg:1967tq}
S.~Weinberg, {\it {A Model of Leptons}},  Phys. Rev. Lett. {\bf 19}
  (1967) 1264.

\bibitem{Salam:1968rm}
A.~Salam, {\it {Weak and Electromagnetic Interactions}},   Conf. Proc. C
  {\bf 680519} (1968) 367.

\bibitem{Higgs:1964pj}
P.~W. Higgs, {\it {Broken Symmetries and the Masses of Gauge Bosons}},  
  Phys. Rev. Lett.  {\bf 13} (1964) 508.

\bibitem{Englert:1964et}
F.~Englert and R.~Brout, {\it {Broken Symmetry and the Mass of Gauge Vector
  Mesons}},  Phys. Rev. Lett.  {\bf 13} (1964) 321.

\bibitem{ATLAS:2018mme}
{\bf ATLAS} Collaboration, M.~Aaboud et~al., {\it {Observation of Higgs boson
  production in association with a top quark pair at the LHC with the ATLAS
  detector}},   Phys. Lett.  {\bf  B 784} (2018) 173
  [\href{http://arxiv.org/abs/1806.00425}{{\tt arXiv:1806.00425}}].

\bibitem{CMS:2018uxb}
{\bf CMS} Collaboration, A.~M. Sirunyan et~al., {\it {Observation of
  $t\bar{t}H$ production}},   Phys. Rev. Lett.  {\bf 120}
  (2018), no.~23, 231801  [\href{http://arxiv.org/abs/1804.02610}{{\tt
  arXiv:1804.02610}}].

\bibitem{ATLAS:2021qou}
{\bf ATLAS} Collaboration, G.~Aad et~al., {\it {Measurement of Higgs boson
  decay into $b$-quarks in associated production with a top-quark pair in $pp$
  collisions at $\sqrt{s}=13$ TeV with the ATLAS detector}},
  JHEP {\bf 06} (2022) 097
  [\href{http://arxiv.org/abs/2111.06712}{{\tt arXiv:2111.06712}}].

\bibitem{ATLAS:2016neq}
{\bf ATLAS, CMS} Collaboration, G.~Aad et~al., {\it {Measurements of the Higgs
  boson production and decay rates and constraints on its couplings from a
  combined ATLAS and CMS analysis of the LHC pp collision data at $ \sqrt{s}=7
  $ and 8 TeV}},   JHEP  {\bf 08} (2016) 045 
  [\href{http://arxiv.org/abs/1606.02266}{{\tt arXiv:1606.02266}}].

\bibitem{ATLAS:2021upq}
{\bf ATLAS} Collaboration, G.~Aad et~al., {\it {Search for charged Higgs bosons
  decaying into a top quark and a bottom quark at $ \sqrt{\mathrm{s}} $ = 13
  TeV with the ATLAS detector}},  JHEP {\bf 06} (2021) 145
  [\href{http://arxiv.org/abs/2102.10076}{{\tt arXiv:2102.10076}}].

\bibitem{CMS:2019rlz}
{\bf CMS} Collaboration, A.~M. Sirunyan et~al., {\it {Search for a charged
  Higgs boson decaying into top and bottom quarks in events with electrons or
  muons in proton-proton collisions at $ \sqrt{\mathrm{s}} $ = 13 TeV}},  
  JHEP  {\bf 01} (2020) 096  [\href{http://arxiv.org/abs/1908.09206}{{\tt
  arXiv:1908.09206}}].

\bibitem{Bredenstein:2008zb}
A.~Bredenstein, A.~Denner, S.~Dittmaier, and S.~Pozzorini, {\it {NLO QCD
  corrections to $t\bar{t}b\bar{b}$ production at the LHC: 1. Quark-antiquark
  annihilation}},  JHEP {\bf 08} (2008) 108 
  [\href{http://arxiv.org/abs/0807.1248}{{\tt arXiv:0807.1248}}].

\bibitem{Bredenstein:2009aj}
A.~Bredenstein, A.~Denner, S.~Dittmaier, and S.~Pozzorini, {\it {NLO QCD
  corrections to $pp\to t\bar{t}b\bar{b}+X$ at the LHC}},   Phys. Rev.
  Lett.  {\bf 103} (2009) 012002  [\href{http://arxiv.org/abs/0905.0110}{{\tt
  arXiv:0905.0110}}].

\bibitem{Bevilacqua:2009zn}
G.~Bevilacqua, M.~Czakon, C.~G. Papadopoulos, R.~Pittau, and M.~Worek, {\it
  {Assault on the NLO Wishlist: $pp\to t\bar{t}b\bar{b}$}},   JHEP  {\bf
  09} (2009) 109  [\href{http://arxiv.org/abs/0907.4723}{{\tt
  arXiv:0907.4723}}].

\bibitem{Bredenstein:2010rs}
A.~Bredenstein, A.~Denner, S.~Dittmaier, and S.~Pozzorini, {\it {NLO QCD
  Corrections to Top Anti-Top Bottom Anti-Bottom Production at the LHC: 2. full
  hadronic results}},   JHEP  {\bf 03} (2010) 021 
  [\href{http://arxiv.org/abs/1001.4006}{{\tt arXiv:1001.4006}}].

\bibitem{Worek:2011rd}
M.~Worek, {\it {On the next-to-leading order QCD $K$-factor for top $t\bar
  tb\bar b$ production at the TeVatron}},   JHEP  {\bf 02} (2012) 043
  [\href{http://arxiv.org/abs/1112.4325}{{\tt arXiv:1112.4325}}].

\bibitem{Bevilacqua:2014qfa}
G.~Bevilacqua and M.~Worek, {\it {On the ratio of $ t\overline{t} b\overline{b}
  $ and $ t\overline{t} jj $ cross sections at the CERN Large Hadron
  Collider}},   JHEP  {\bf 07} (2014) 135
  [\href{http://arxiv.org/abs/1403.2046}{{\tt arXiv:1403.2046}}].

\bibitem{Bevilacqua:2018dny}
G.~Bevilacqua, H.~B. Hartanto, M.~Kraus, T.~Weber, and M.~Worek, {\it {Precise
  predictions for $t\bar{t}\gamma/t\bar{t}$ cross section ratios at the LHC}},
  JHEP  {\bf 01} (2019) 188 [\href{http://arxiv.org/abs/1809.08562}{{\tt
  arXiv:1809.08562}}].

\bibitem{Bevilacqua:2020srb}
G.~Bevilacqua, H.-Y. Bi, H.~B. Hartanto, M.~Kraus, J.~Nasufi, and M.~Worek,
  {\it {NLO QCD corrections to off-shell ${t\bar{t}W^\pm}$ production at the
  LHC: Correlations and Asymmetries}},   Eur. Phys. J.  {\bf C 81} (2021)
  675 [\href{http://arxiv.org/abs/2012.01363}{{\tt arXiv:2012.01363}}].

\bibitem{Buccioni:2019plc}
F.~Buccioni, S.~Kallweit, S.~Pozzorini, and M.~F. Zoller, {\it {NLO QCD
  predictions for $t\bar{t}b\bar{b}$ production in association with a light jet
  at the LHC}},   JHEP {\bf 12} (2019) 015
  [\href{http://arxiv.org/abs/1907.13624}{{\tt arXiv:1907.13624}}].

\bibitem{Kardos:2013vxa}
A.~Kardos and Z.~Tr\'ocs\'anyi, {\it {Hadroproduction of t anti-t pair with a b
  anti-b pair using PowHel}},  J. Phys. {\bf G 41} (2014) 075005 
  [\href{http://arxiv.org/abs/1303.6291}{{\tt arXiv:1303.6291}}].

\bibitem{Garzelli:2014aba}
M.~V. Garzelli, A.~Kardos, and Z.~Tr\'ocs\'anyi, {\it {Hadroproduction of
  $t\bar{t}b\bar{b}$ final states at LHC: predictions at NLO accuracy matched
  with Parton Shower}},   JHEP  {\bf 03} (2015) 083 
  [\href{http://arxiv.org/abs/1408.0266}{{\tt arXiv:1408.0266}}].

\bibitem{Bevilacqua:2017cru}
G.~Bevilacqua, M.~V. Garzelli, and A.~Kardos, {\it {$t\bar{t}b\bar{b}$
  hadroproduction with massive bottom quarks with PowHel}},
  \href{http://arxiv.org/abs/1709.06915}{{\tt arXiv:1709.06915}}.

\bibitem{Jezo:2018yaf}
T.~Je\v{z}o, J.~M. Lindert, N.~Moretti, and S.~Pozzorini, {\it {New NLOPS
  predictions for $t \bar{t} +b$ -jet production at the LHC}},  Eur. Phys.
  J.  {\bf C 78} (2018), no.~6, 502
  [\href{http://arxiv.org/abs/1802.00426}{{\tt arXiv:1802.00426}}].

\bibitem{Cascioli:2013era}
F.~Cascioli, P.~Maierh\"ofer, N.~Moretti, S.~Pozzorini, and F.~Siegert, {\it
  {NLO matching for $t\bar t b \bar b$ production with massive $b$-quarks}},
  Phys. Lett.  {\bf B 734} (2014) 210
  [\href{http://arxiv.org/abs/1309.5912}{{\tt arXiv:1309.5912}}].

\bibitem{CMS:2013vui}
{\bf CMS Collaboration} Collaboration, {\it {Measurement of the cross section
  ratio $t\bar{t}b\bar{b}/t\bar{t}jj$ in pp Collisions at 8 TeV}},  tech. rep.,
  CERN (2013) CMS-PAS-TOP-13-010.

\bibitem{ATLAS:2018fwl}
{\bf ATLAS} Collaboration, M.~Aaboud et~al., {\it {Measurements of inclusive
  and differential fiducial cross-sections of $ t\overline{t} $ production with
  additional heavy-flavour jets in proton-proton collisions at $ \sqrt{s} $ =
  13 TeV with the ATLAS detector}},   JHEP {\bf 04} (2019) 046 
  [\href{http://arxiv.org/abs/1811.12113}{{\tt arXiv:1811.12113}}].

\bibitem{CMS:2017xnm}
{\bf CMS} Collaboration, A.~M. Sirunyan et~al., {\it {Measurements of
  $t\bar{t}$ cross sections in association with $b$ jets and inclusive jets and
  their ratio using dilepton final states in pp collisions at $\sqrt{s}$ = 13
  TeV}},  Phys. Lett. {\bf B 776} (2018) 355
  [\href{http://arxiv.org/abs/1705.10141}{{\tt arXiv:1705.10141}}].

\bibitem{CMS:2020grm}
{\bf CMS} Collaboration, A.~M. Sirunyan et~al., {\it {Measurement of the cross
  section for $\text{t}\bar{\text{t}}$ production with additional jets and b
  jets in pp collisions at $\sqrt{s}=$ 13 TeV}},  JHEP  {\bf 07} (2020)
  125 [\href{http://arxiv.org/abs/2003.06467}{{\tt arXiv:2003.06467}}].

\bibitem{Denner:2020orv}
A.~Denner, J.-N. Lang, and M.~Pellen, {\it {Full NLO QCD corrections to
  off-shell $t\bar{t}b\bar{b}$ production}},  Phys. Rev. {\bf
D 104}  (2021), no.~5, 056018 [\href{http://arxiv.org/abs/2008.00918}{{\tt
  arXiv:2008.00918}}].

\bibitem{Bevilacqua:2021cit}
G.~Bevilacqua, H.-Y. Bi, H.~B. Hartanto, M.~Kraus, M.~Lupattelli, and M.~Worek,
  {\it {$t\bar{t}b\bar{b}$ at the LHC: on the size of corrections
  and b-jet definitions}},   JHEP {\bf 08} (2021) 008
  [\href{http://arxiv.org/abs/2105.08404}{{\tt arXiv:2105.08404}}].

\bibitem{Denner:2014wka}
A.~Denner, R.~Feger, and A.~Scharf, {\it {Irreducible background and
  interference effects for Higgs-boson production in association with a
  top-quark pair}},   JHEP {\bf 04} (2015) 008
  [\href{http://arxiv.org/abs/1412.5290}{{\tt arXiv:1412.5290}}].

\bibitem{Bevilacqua:2019quz}
G.~Bevilacqua, H.~B. Hartanto, M.~Kraus, T.~Weber, and M.~Worek, {\it
  {Off-shell vs on-shell modelling of top quarks in photon associated
  production}},   JHEP {\bf 03} (2020) 154
  [\href{http://arxiv.org/abs/1912.09999}{{\tt arXiv:1912.09999}}].

\bibitem{Bevilacqua:2011xh}
G.~Bevilacqua, M.~Czakon, M.~V. Garzelli, A.~van Hameren, A.~Kardos, C.~G.
  Papadopoulos, R.~Pittau, and M.~Worek, {\it {HELAC-NLO}},  Comput. Phys.
  Commun.  {\bf 184} (2013) 986
  [\href{http://arxiv.org/abs/1110.1499}{{\tt arXiv:1110.1499}}].

\bibitem{Cafarella:2007pc}
A.~Cafarella, C.~G. Papadopoulos, and M.~Worek, {\it {Helac-Phegas: A Generator
  for all parton level processes}},  Comput. Phys. Commun.  {\bf 180}
  (2009) 1941 [\href{http://arxiv.org/abs/0710.2427}{{\tt
  arXiv:0710.2427}}].

\bibitem{vanHameren:2007pt}
A.~van Hameren, {\it {PARNI for importance sampling and density estimation}},
   Acta Phys. Polon.  {\bf B 40} (2009) 259 
  [\href{http://arxiv.org/abs/0710.2448}{{\tt arXiv:0710.2448}}].

\bibitem{vanHameren:2010gg}
A.~van Hameren, {\it {Kaleu: A General-Purpose Parton-Level Phase Space
  Generator}},  \href{http://arxiv.org/abs/1003.4953}{{\tt arXiv:1003.4953}}.

\bibitem{vanHameren:2009dr}
A.~van Hameren, C.~G. Papadopoulos, and R.~Pittau, {\it {Automated one-loop
  calculations: A Proof of concept}},   JHEP  {\bf 09} (2009) 106
  [\href{http://arxiv.org/abs/0903.4665}{{\tt arXiv:0903.4665}}].

\bibitem{Ossola:2006us}
G.~Ossola, C.~G. Papadopoulos, and R.~Pittau, {\it {Reducing full one-loop
  amplitudes to scalar integrals at the integrand level}},   Nucl. Phys. B
  {\bf B 763} (2007) 147  [\href{http://arxiv.org/abs/hep-ph/0609007}{{\tt
  hep-ph/0609007}}].

\bibitem{Ossola:2007ax}
G.~Ossola, C.~G. Papadopoulos, and R.~Pittau, {\it {CutTools: A Program
  implementing the OPP reduction method to compute one-loop amplitudes}},  
  JHEP {\bf 03} (2008) 042 [\href{http://arxiv.org/abs/0711.3596}{{\tt
  arXiv:0711.3596}}].

\bibitem{vanHameren:2010cp}
A.~van Hameren, {\it {OneLOop: For the evaluation of one-loop scalar
  functions}},  Comput. Phys. Commun. {\bf 182} (2011) 2427 
  [\href{http://arxiv.org/abs/1007.4716}{{\tt arXiv:1007.4716}}].

\bibitem{Czakon:2009ss}
M.~Czakon, C.~G. Papadopoulos, and M.~Worek, {\it {Polarizing the Dipoles}},
  JHEP {\bf 08} (2009) 085  [\href{http://arxiv.org/abs/0905.0883}{{\tt
  arXiv:0905.0883}}].

\bibitem{Catani:1996vz}
S.~Catani and M.~H. Seymour, {\it {A General algorithm for calculating jet
  cross-sections in NLO QCD}},   Nucl. Phys.  {\bf B 485} (1997) 291,
  [\href{http://arxiv.org/abs/hep-ph/9605323}{{\tt hep-ph/9605323}}]. [Erratum:
  Nucl. Phys. {\bf B 510}   (1998) 503].

\bibitem{Catani:2002hc}
S.~Catani, S.~Dittmaier, M.~H. Seymour, and Z.~Trocsanyi, {\it {The Dipole
  formalism for next-to-leading order QCD calculations with massive partons}},
   Nucl. Phys.   {\bf B 627} (2002) 189 
  [\href{http://arxiv.org/abs/hep-ph/0201036}{{\tt hep-ph/0201036}}].

\bibitem{Bevilacqua:2013iha}
G.~Bevilacqua, M.~Czakon, M.~Kubocz, and M.~Worek, {\it {Complete Nagy-Soper
  subtraction for next-to-leading order calculations in QCD}},   JHEP  {\bf
  10} (2013) 204  [\href{http://arxiv.org/abs/1308.5605}{{\tt
  arXiv:1308.5605}}].

\bibitem{Campbell:2004ch}
J.~M. Campbell, R.~K. Ellis, and F.~Tramontano, {\it {Single top production and
  decay at next-to-leading order}},  Phys. Rev.  {\bf  D 70} (2004)
094012 
  [\href{http://arxiv.org/abs/hep-ph/0408158}{{\tt hep-ph/0408158}}].

\bibitem{Nagy:1998bb}
Z.~Nagy and Z.~Trocsanyi, {\it {Next-to-leading order calculation of four jet
  observables in electron positron annihilation}},   Phys. Rev. 
{\bf D 59}
  (1999) 014020  [\href{http://arxiv.org/abs/hep-ph/9806317}{{\tt
  hep-ph/9806317}}]  [Erratum: Phys. Rev. {\bf D 62} (2000) 099902].

\bibitem{Nagy:2003tz}
Z.~Nagy, {\it {Next-to-leading order calculation of three jet observables in
  hadron hadron collision}},  Phys. Rev.  {\bf D 68} (2003) 094002 
  [\href{http://arxiv.org/abs/hep-ph/0307268}{{\tt hep-ph/0307268}}].

\bibitem{Czakon:2015cla}
M.~Czakon, H.~B. Hartanto, M.~Kraus, and M.~Worek, {\it {Matching the
  Nagy-Soper parton shower at next-to-leading order}},   JHEP  {\bf 06}
  (2015) 033  [\href{http://arxiv.org/abs/1502.00925}{{\tt arXiv:1502.00925}}].

\bibitem{Alwall:2006yp}
J.~Alwall et~al., {\it {A Standard format for Les Houches event files}},  
  Comput. Phys. Commun.  {\bf 176} (2007) 300 
  [\href{http://arxiv.org/abs/hep-ph/0609017}{{\tt hep-ph/0609017}}].

\bibitem{Antcheva:2009zz}
I.~Antcheva et~al., {\it {ROOT: A C++ framework for petabyte data storage,
  statistical analysis and visualization}},  Comput. Phys. Commun.  {\bf
  180} (2009) 2499  [\href{http://arxiv.org/abs/1508.07749}{{\tt
  arXiv:1508.07749}}].

\bibitem{Denner:1999gp}
A.~Denner, S.~Dittmaier, M.~Roth and D.~Wackeroth,
{\it Predictions for all processes $e^+ e^- + 4$ fermions
  $+\,\gamma$},  Nucl. Phys.  \textbf{B 560} (1999)  33
 [\href{https://arxiv.org/abs/hep-ph/9904472}{{\tt
  arXiv:hep-ph/9904472}}]

\bibitem{Denner:2005fg}
A.~Denner, S.~Dittmaier, M.~Roth and L.~H.~Wieders,
{\it Electroweak corrections to charged-current $e^+ e^- \to 4$
  fermion processes: Technical details and further results},
  Nucl. Phys.   \textbf{B 724} (2005)  247
 [Erratum:    Nucl. Phys.  \textbf{ B 854} (2012)  504]
[\href{https://arxiv.org/abs/hep-ph/0505042}{{\tt
  arXiv:hep-ph/0505042}}].
 
\bibitem{Frederix:2018nkq} R.~Frederix, S.~Frixione, V.~Hirschi, D.~Pagani,
H.~S.~Shao and M.~Zaro, {\it {The automation of next-to-leading order
    electroweak
calculations}},  JHEP  \textbf{07} (2018)  185 [Erratum: 
JHEP  \textbf{11} (2021)  085]
[\href{https://arxiv.org/abs/1804.10017}{{\tt arXiv:1804.10017}}].


\bibitem{Jezabek:1988iv}
M.~Je\.zabek and J.~H. K\"uhn, {\it {QCD Corrections to Semileptonic Decays of
  Heavy Quarks}},  Nucl. Phys. {\bf B 314} (1989) 1.

\bibitem{Basso:2015gca}
L.~Basso, S.~Dittmaier, A.~Huss, and L.~Oggero, {\it {Techniques for the
  treatment of IR divergences in decay processes at NLO and application to the
  top-quark decay}},  Eur. Phys. J.  {\bf  C 76} (2016), no.~2, 56
  [\href{http://arxiv.org/abs/1507.04676}{{\tt arXiv:1507.04676}}].

\bibitem{Denner:2012yc}
A.~Denner, S.~Dittmaier, S.~Kallweit, and S.~Pozzorini, {\it {NLO QCD
  corrections to off-shell top-antitop production with leptonic decays at
  hadron colliders}},   JHEP  {\bf 10} (2012) 110 
  [\href{http://arxiv.org/abs/1207.5018}{{\tt arXiv:1207.5018}}].

\bibitem{Buckley:2014ana}
A.~Buckley, J.~Ferrando, S.~Lloyd, K.~Nordstr\"om, B.~Page, M.~R\"ufenacht,
  M.~Sch\"onherr, and G.~Watt, {\it {LHAPDF6: parton density access in the LHC
  precision era}},   Eur. Phys. J. {\bf C 75} (2015) 132 
  [\href{http://arxiv.org/abs/1412.7420}{{\tt arXiv:1412.7420}}].

\bibitem{NNPDF:2017mvq}
{\bf NNPDF} Collaboration, R.~D. Ball et~al., {\it {Parton distributions from
  high-precision collider data}},  Eur. Phys. J.  {\bf C 77} (2017)
  no.~10, 663  [\href{http://arxiv.org/abs/1706.00428}{{\tt arXiv:1706.00428}}].

\bibitem{Cacciari:2008gp}
M.~Cacciari, G.~P. Salam, and G.~Soyez, {\it {The anti-$k_t$ jet clustering
  algorithm}},   JHEP  {\bf 04} (2008) 063
  [\href{http://arxiv.org/abs/0802.1189}{{\tt arXiv:0802.1189}}].

  \bibitem{Melnikov:2009dn}
K.~Melnikov and M.~Schulze, {\it {NLO QCD corrections to top quark pair
  production and decay at hadron colliders}},  JHEP  {\bf 08} (2009)
049 
  [\href{http://arxiv.org/abs/0907.3090}{{\tt arXiv:0907.3090}}].

\bibitem{Campbell:2012uf}
J.~Campbell and K.~Ellis, {\it Top-Quark Processes at NLO in Production 
and Decay}, J.Phys.G 42 (2015) 1, 015005 
[\href{https://arxiv.org/abs/1204.1513}{{\tt arXiv:1204.1513}}].

\bibitem{Melnikov:2011ta}
K.~Melnikov, M.~Schulze, and A.~Scharf, {\it {QCD corrections to top quark pair
  production in association with a photon at hadron colliders}},   Phys.
  Rev.   {\bf D 83} (2011) 074013  [\href{http://arxiv.org/abs/1102.1967}{{\tt
  arXiv:1102.1967}}].

\bibitem{Melnikov:2011qx}
K.~Melnikov, A.~Scharf, and M.~Schulze, {\it {Top quark pair production in
  association with a jet: QCD corrections and jet radiation in top quark
  decays}},   Phys. Rev. {\bf D 85} (2012) 054002 
  [\href{http://arxiv.org/abs/1111.4991}{{\tt arXiv:1111.4991}}].

\bibitem{Behring:2019iiv}
A.~Behring, M.~Czakon, A.~Mitov, A.~S.~Papanastasiou and R.~Poncelet, {\it Higher order corrections to spin correlations in top quark pair production at the LHC}, Phys. Rev. Lett. \textbf{123} (2019) no.8, 082001 
[\href{https://arxiv.org/abs/1901.05407}{{\tt arXiv:1901.05407}}].

\bibitem{Czakon:2020qbd}
M.~Czakon, A.~Mitov, and R.~Poncelet, {\it {NNLO QCD corrections to leptonic
  observables in top-quark pair production and decay}},   JHEP   {\bf 05}
  (2021) 212 [\href{http://arxiv.org/abs/2008.11133}{{\tt arXiv:2008.11133}}].

\bibitem{Denner:2014zga}
A.~Denner and J.~N.~Lang,
{\it The Complex-Mass Scheme and Unitarity in perturbative Quantum
  Field Theory},
Eur. Phys. J. {\bf C 75} (2015) no.8, 377
[\href{http://arxiv.org/abs/1406.6280}{{\tt arXiv:1406.6280}}].

\bibitem{Denner:2006ic}
A.~Denner and S.~Dittmaier,
{\it The Complex-mass scheme for perturbative calculations with
  unstable particles},
Nucl. Phys. B Proc. Suppl.  {\bf 160} (2006) 22
[\href{https://arxiv.org/abs/hep-ph/0605312}{{\tt hep-ph/0605312}}].

\bibitem{SM:2012sed}
{\bf SM, NLO MULTILEG Working Group, SM MC Working Group} Collaboration,
  J.~Alcaraz~Maestre et~al., {\it {The SM and NLO Multileg and SM MC Working
  Groups: Summary Report}},  in 7th Les Houches Workshop on Physics at
  TeV Colliders, pp.~1--220, 3, 2012.
\newblock \href{http://arxiv.org/abs/1203.6803}{{\tt arXiv:1203.6803}}.

\bibitem{Bevilacqua:2020pzy}
G.~Bevilacqua, H.-Y. Bi, H.~B. Hartanto, M.~Kraus, and M.~Worek, {\it {The
  simplest of them all: $t\bar{t} W^\pm$ at NLO accuracy in QCD}},   JHEP
  {\bf 08} (2020) 043  [\href{http://arxiv.org/abs/2005.09427}{{\tt
  arXiv:2005.09427}}].

\bibitem{Hermann:2021xvs}
J.~Hermann and M.~Worek, {\it {The impact of top-quark modelling on the
  exclusion limits in ${t\bar{t}}+{DM}$ searches at the LHC}},   Eur. Phys.
  J.  {\bf C 81} (2021)  no.~11, 1029
  [\href{http://arxiv.org/abs/2108.01089}{{\tt arXiv:2108.01089}}].

\bibitem{Bevilacqua:2021tzp}
G.~Bevilacqua, H.~Y. Bi, F.~Febres~Cordero, H.~B. Hartanto, M.~Kraus,
  J.~Nasufi, L.~Reina, and M.~Worek, {\it {Modeling uncertainties of
  $t\bar{t}W^\pm$ multilepton signatures}},  Phys. Rev. {\bf D
105}
  (2022)  no.~1, 014018  [\href{http://arxiv.org/abs/2109.15181}{{\tt
  arXiv:2109.15181}}].

\bibitem{Stremmer:2021bnk}
D.~Stremmer and M.~Worek, {\it {Production and decay of the Higgs boson in
    association with top quarks}},  JHEP {\bf 02} (2022) 196
[\href{http://arxiv.org/abs/2111.01427}{{\tt  arXiv:2111.01427}}].

\bibitem{Bevilacqua:2010qb}
G.~Bevilacqua, M.~Czakon, A.~van Hameren, C.~G.~Papadopoulos and
M.~Worek, {\it Complete off-shell effects in top quark pair
  hadroproduction with leptonic decay at next-to-leading order}, 
JHEP {\bf 02} (2011) 083 
[\href{https://arxiv.org/abs/1012.4230}{\tt arXiv:1012.4230}].

\bibitem{Kauer:2001sp}
N.~Kauer and D.~Zeppenfeld, {\it {Finite width effects in top quark production
  at hadron colliders}},   Phys. Rev.  {\bf D 65} (2002) 014021
  [\href{http://arxiv.org/abs/hep-ph/0107181}{{\tt hep-ph/0107181}}].

\bibitem{Liebler:2015ipp}
S.~Liebler, G.~Moortgat-Pick, and A.~S. Papanastasiou, {\it {Probing the
  top-quark width through ratios of resonance contributions of
  $e^+e^-\rightarrow W^+W^-b\bar{b}$}},   JHEP  {\bf 03} (2016) 099
  [\href{http://arxiv.org/abs/1511.02350}{{\tt arXiv:1511.02350}}].

\bibitem{Baskakov:2018huw}
A.~Baskakov, E.~Boos, and L.~Dudko, {\it {Model independent top quark width
  measurement using a combination of resonant and nonresonant cross sections}},
    Phys. Rev.  {\bf D 98} (2018), no.~11, 116011
  [\href{http://arxiv.org/abs/1807.11193}{{\tt arXiv:1807.11193}}].

\bibitem{ATLAS:2016muw}
{\bf ATLAS} Collaboration, M.~Aaboud et~al., {\it {Measurement of the top quark
  mass in the $t\bar{t}\to$ dilepton channel from $\sqrt{s}=8$ TeV ATLAS
  data}},  Phys. Lett.   {\bf B 761} (2016) 350
  [\href{http://arxiv.org/abs/1606.02179}{{\tt arXiv:1606.02179}}].

\bibitem{CMS:2017znf}
{\bf CMS} Collaboration, A.~M. Sirunyan et~al., {\it {Measurement of the top
  quark mass in the dileptonic $t\bar{t}$ decay channel using the mass
  observables $M_{b\ell}$, $M_{T2}$, and $M_{b\ell\nu}$ in pp collisions at
  $\sqrt{s}=8$ TeV}},   Phys. Rev.  {\bf D 96} (2017), no.~3, 032002
  [\href{http://arxiv.org/abs/1704.06142}{{\tt arXiv:1704.06142}}].

\bibitem{Haisch:2018djm}
U.~Haisch and G.~Polesello, {\it {Searching for heavy Higgs bosons in the $t
  \bar t Z$ and $t b W$ final states}},   JHEP {\bf 09} (2018) 151
  [\href{http://arxiv.org/abs/1807.07734}{{\tt arXiv:1807.07734}}].

\bibitem{Haisch:2018bby}
U.~Haisch and G.~Polesello, {\it {Searching for production of dark matter in
  association with top quarks at the LHC}},  JHEP  {\bf 02} (2019) 029
  [\href{http://arxiv.org/abs/1812.00694}{{\tt arXiv:1812.00694}}].

\bibitem{CMS:2018ncu}
{\bf CMS} Collaboration, A.~M. Sirunyan et~al., {\it {Search for pair
  production of first-generation scalar leptoquarks at $\sqrt{s} =$ 13 TeV}},
   Phys. Rev. {\bf D 99} (2019), no.~5, 052002
  [\href{http://arxiv.org/abs/1811.01197}{{\tt arXiv:1811.01197}}].

  \bibitem{Lester:1999tx}
C.~G. Lester and D.~J. Summers, {\it {Measuring masses of semiinvisibly
  decaying particles pair produced at hadron colliders}},   Phys. Lett. 
  {\bf B 463} (1999) 99  [\href{http://arxiv.org/abs/hep-ph/9906349}{{\tt
  hep-ph/9906349}}].

\bibitem{Barr:2003rg}
A.~Barr, C.~Lester, and P.~Stephens, {\it {m(T2): The Truth behind the
  glamour}},   J. Phys. {\bf  G 29} (2003) 2343
  [\href{http://arxiv.org/abs/hep-ph/0304226}{{\tt hep-ph/0304226}}].

\bibitem{Lester:2014yga}
C.~G. Lester and B.~Nachman, {\it {Bisection-based asymmetric M$_{T2}$
  computation: a higher precision calculator than existing symmetric methods}},
    JHEP {\bf 03} (2015) 100 [\href{http://arxiv.org/abs/1411.4312}{{\tt
  arXiv:1411.4312}}].

\bibitem{ATLAS:2014gmw}
{\bf ATLAS} Collaboration, G.~Aad et~al., {\it {Search for direct top-squark
  pair production in final states with two leptons in pp collisions at
  $\sqrt{s} =$ 8TeV with the ATLAS detector}},   JHEP  {\bf 06} (2014)
124 
  [\href{http://arxiv.org/abs/1403.4853}{{\tt arXiv:1403.4853}}].

\bibitem{Haisch:2016gry}
U.~Haisch, P.~Pani, and G.~Polesello, {\it {Determining the CP nature of spin-0
  mediators in associated production of dark matter and $ t\overline{t} $
  pairs}},  JHEP {\bf 02} (2017) 131 
  [\href{http://arxiv.org/abs/1611.09841}{{\tt arXiv:1611.09841}}].

\bibitem{Choi:2019lyt}
J.~Choi, T.~J. Kim, J.~Lim, J.~Park, Y.~Ryou, J.~Song, and S.~Yun, {\it
  {Identification of Additional Jets in the $t\bar{t}b\bar{b}$
  Events by Using Deep Neural Network}},   J. Korean Phys. Soc.  {\bf 77}
  (2020), no.~12, 1100 [\href{http://arxiv.org/abs/1910.14535}{{\tt
  arXiv:1910.14535}}].

\bibitem{Jang:2021eph}
C.~Jang, S.-K. Ko, Y.-K. Noh, J.~Choi, J.~Lim, and T.~J. Kim, {\it {Learning to
  increase matching efficiency in identifying additional b-jets in the
  $\text{t}\bar{\text{t}}\text{b}\bar{\text{b}}$ process}},
  [\href{http://arxiv.org/abs/2103.09129}{{\tt arXiv:2103.09129}}].

\bibitem{Cavallini:2021vot}
L.~Cavallini, A.~Coccaro, C.~K. Khosa, G.~Manco, S.~Marzani, F.~Parodi,
  D.~Rebuzzi, A.~Rescia, and G.~Stagnitto, {\it {Tagging the Higgs boson decay
  to bottom quarks with colour-sensitive observables and the Lund jet
  plane}},
  Eur. Phys. J. {\bf C 82} (2022) 5, 493
  [\href{http://arxiv.org/abs/2112.09650}{{\tt arXiv:2112.09650}}].

\bibitem{LHCHiggsCrossSectionWorkingGroup:2016ypw}
{\bf LHC Higgs Cross Section Working Group} Collaboration, D.~de~Florian
  et~al., {\it {Handbook of LHC Higgs Cross Sections: 4. Deciphering the Nature
  of the Higgs Sector}},   CERN Yellow Reports: Monographs  {\bf 2/2017}
  (2016) [\href{http://arxiv.org/abs/1610.07922}{{\tt arXiv:1610.07922}}].

\end{thebibliography}
\end{document}